# A Universal Polybromide Melt Strategy for the Direct Conversion of Metals into Optoelectronic-Grade Bromide Perovskites

Arindam Mondal[1]*, Mark Baranov[2], Lonia R. Friedlander[2], Mariela J. Pavan[2], Kirill Sobolev[1], Yevgeny Rakita[1,2]*

[1] Department of Materials Engineering, Ben Gurion University of the Negev, Be'er Sheva, Israel

[2] Ilse Katz Institute for Nanoscale Science and Technology, Ben-Gurion University of the Negev, Be'er Sheva, Israel

*Corresponding Authors: rakita@bgu.ac.il, arindamm@post.bgu.ac.il

## Abstract

Halide perovskite systems require specific metal oxidation states ($M^{2+}$ or $M^{+}/M^{3+}$ pairs), limiting processing flexibility during traditional synthesis and posing stability challenges for multivalent metals like Sn, Cu, and Au. This work anchors a profound paradigm shift: a universal, single-step, organic-solvent-free synthetic platform that directly converts stable elemental metal films into bromide perovskite and perovskite-related systems for a wide variety (opto-)electronic applications. The synthetic platform uses methylammonium polybromide melt (PBM) that reacts at room temperature with different elemental metals, including Pb, Sn, Au, and Cu, successfully transforming them into their respective Br-perovskite phases: $MAPbBr_3$, $MASnBr_3$, $MA_2Au_2Br_6$, and $MA_2CuBr_4$. It is shown that the PBM's chemical activity can be tuned, and *in situ* X-ray diffraction captures in two cases transient intermediate phases during phase evolution. For Sn, which tends towards overoxidation to $Sn^{4+}$ phase, changes in the PBM's composition coupled with a metallic-Sn-mediated comproportionation reaction allowed suppressing overoxidation, and formation of high-quality $MASnBr_3$ films under ambient conditions. Overall, oxidation and disproportionation potentials are found to be useful metrics for predicting M--PBM reaction outcomes. Industrially, this approach should be fully compatible with roll-to-roll infrastructure, providing a low-barrier pathway to scale next-generation optoelectronic technologies.

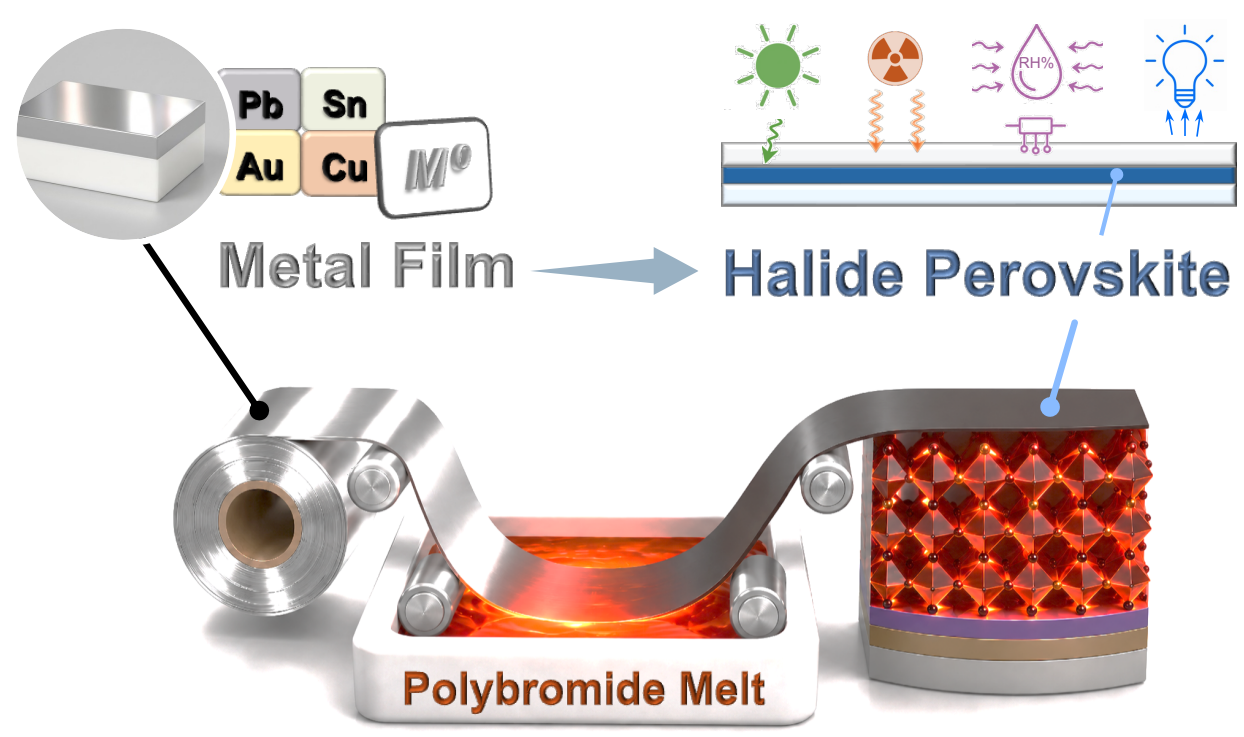

**Summary**: A novel, universal polybromide-melt strategy is reported, allowing a single-step conversion of elemental metal films (e.g., Pb, Sn, Cu, and Au) into methylammonium bromide perovskite phases, for next-generation optoelectronic devices. The investigated chemistry provides a deeper understanding of the redox chemistry and stability of halide perovskites and offers a scalable route with broad applicability across the optoelectronic materials domain.

## 1. Introduction

Halide perovskites (HaPs) are a class of optoelectronic materials, best known for their exceptional potential as absorbers for solar energy, achieving light-to-electron conversion efficiencies of 26.9% as a single junction or 35.0% in tandem with Si.[1] Besides solar-cells, HaPs are relevant to many other optoelectronic applications, generating efficient light-emitting diodes, photodetectors for high-energy radiation, lasers, humidity sensors, photocatalysts, memristors, and more.[2–4] In most cases, their fundamental physical properties, such as high absorption coefficients, low exciton binding energies, self-healing, low (active) defect density, and long charge carrier diffusion lengths, [5–11] provide the foundation for their superior functionality.

HaP synthesis is often solution-based – much simpler than classical semiconductors such as Si and GaAs – making them very attractive for industry, especially as a future source of materials to meet the growing demand for renewable energy.[12,13] Therefore, to make the HaP synthesis and operation robust and reproducible, the production chemistry and operation of the perovskite system must be well understood and strictly controlled. [2,14–16]

The correct oxidation state of the atoms involved in HaP(-related) structures is a primary requirement for their formation. Generally, HaPs carry an $AMX_3$ stoichiometry, where: '*A*' is a *monovalent cation* like $Cs^+$, formamidinium ($FA^+$) or methylammonium ($MA^+$); '*X*' is a *monovalent anion*, like $Br^-$ or $I^-$; '*M*' is a metal-group *divalent cation*. Unlike *A* and *X* groups, *M* may show some flexibility in its oxidation and still form perovskite-related structures. While often kept at 2+, such as $Pb^{2+}$, $Sn^{2+}$, $Ge^{2+}$, $Cu^{2+}$, $Mn^{2+}$, it can also construct super-cells with alternating $M^{I}$-$M^{III}$ pairs at *M* sites, such as heteroatomic Ag-Bi, Ag-Fe, Cu-Sb, or homoatomic Au-Au.[17] In a broader sense, low-dimensional perovskite-related systems that share the same perovskite-like corner-sharing [$MX_6$] octahedra, such as the 2D $A_2BX_4$, impose similar constraints on the oxidation states of the elements they are composed of. Similar to 3D HaPs, low-dimensional (HaP-related) systems also show interesting optoelectronic, dielectric, and magnetic properties.[18]

Precise control over oxidation states during synthesis and throughout operation is of the utmost importance, where any alteration in the oxidation states of the precursors can lead to undesired reaction products. Since *A* and *X* ions often carry a single/dominant oxidation state, it is most important to have good control over the oxidation state of the *M* cation – a requirement that may be challenging in many cases. In fact, the commonly used $Pb^{2+}$ cation at the *M* site is one of the few cases where the 2+ oxidation state is chemically preferred. Other elements from which HaPs are made, such as Sn, Ge, Au, Cu, Mn, and others, tend to have multiple oxidation states, where the 2+ is not necessarily their most stable.

Tightly related to the oxidation control are the degradation processes happening in HaP-related systems. While *reversible* degradation (self-healing) is known to occur in HaPs,[8] and which is often related to their dynamic structural nature[19,20] and low formation and kinetic energies[11], HaPs are not immune to *irreversible* degradation, which is generally related to electrochemical activity. In HaP photovoltaic devices, degradation pathways are often observed during illumination and/or under electrical bias and usually involve *photolysis* or *electrolysis*. For example, Au-halide-based systems are known to undergo photolysis [21] and disproportionation[22] that can also be induced by an X-ray beam[23]. In these processes, reactive and/or volatile species form, such as halides, halogens or polyhalides, and methylamine gas, which can evaporate or percolate and react with selective charge-transport layers and current-collectors.[24,25] In addition, photolytic and electrolytic processes may lead to the formation of competing phases, often accompanied by changes in the *M* oxidation states.[25,26] Overall, a better understanding and control of (photo)electrochemical processes in HaP systems will enable the development of more robust and rational design rules for HaP-based technologies, thereby increasing their stability and longevity during use.

The constraint on oxidation state is a major weakness in the synthesis of HaPs. Besides the need to control the oxidation state of the desired product, there is often a strict requirement to use a stable precursor with the correct oxidation state. The main question we will be tackling in this paper is whether there is an alternative, generic pathway for oxidizing *M* precursors to the desired oxidation state, specifically *from their elemental metallic, $M^0$, form*.

Utilizing metallic precursors to directly synthesize HaPs can solve multiple challenges in the formation of optoelectronic thin films: toxicity, storage costs, and scalability. Pb, for instance, is mostly toxic in its oxidized $Pb^{2+}$ state and not in its pure metallic state. As for stability maintenance: Sn and Au are required to have $Sn^{2+}$ and $Au^{+}$ states to form functionally relevant perovskite systems. However, $Sn^{2+}$ tends to spontaneously oxidize to $Sn^{4+}$ in ambient atmosphere, and $Au^{+}$ (for $A_2Au^{I}Au^{III}X_6$) tends to spontaneously disproportionate to $Au^0$ and $Au^{3+}$.[27,28] Therefore, manufacturing, storing, and maintaining a highly controlled non-oxidizing reaction environment for such less-stable precursors is clearly more complicated and expensive. Finally, large-scale roll-to-roll metal deposition processes for elemental metals are very mature,[29] so direct conversion of metals into functional perovskites will clearly benefit from it and act as an accelerator.

**Between Metals, Polyhalides, and HaPs:**

Both processes, photolysis and electrolysis, can occur spontaneously during the operation of a solar-cell – the solar light can lead to photolysis, and the built-up voltage can lead to electrolysis. The products of these (photo)electrochemical reactions can be generally formulated as follows:

$$\text{Eq. 1} \quad AMX_3 \xleftrightarrow{h\nu \mid V} M^0 + X_2 + AX \longleftrightarrow M^0 + AX_3$$

The reaction products are a reduced metal and a molecular halogen (=oxidized halide), which often spontaneously react with a reduced halide, e.g., an *AX* salt, forming polyhalides with the general formula $AX_{2n+1}$ ($n$=1,2,…).[30] Methylammonium-based polyhalides, such as $MAI_3$, are liquid at room temperature and highly reactive.[31,32] Due to the liquid nature and high reactivity of the polyhalides, and the liquid-*like* nature of $MAPbI_3$,[19,20] it is not unreasonable to assume that $MAI_3$ may project its characteristics, or even be present in some quantity in $MAPbI_3$. In an active solar cell, meaning under light and bias, the polyhalides may also play an active role in the degradation and self-healing of $MAPbI_3$-based solar-cells[33,34].

The reverse reactions, where metals are spontaneously oxidized and react with halogens or polyhalides, forming metal-halides, are known [35,36], and, in fact, this reverse reaction explain the regenerative pathway induced by photolysis and/or electrolysis degradation [37]. For example, recent work demonstrated that light-induced spontaneous formation of $Pb^0$ metal centers in $CsPbBr_3$ nanoparticles is regenerated in the presence of a polybromide.[37] Earlier works showed that one can convert a metallic Pb film into a HaP using a polyiodide melt or polyhalides dissolved in an alcoholic solution; [31,38] in both cases, an *A* cation (MA or FA) was present as a counter ion for the polyhalide. In the solution process,[38] it was also shown that Sn-HaPs can be formed, but the formed $Sn^{2+}$ perovskite phase was rapidly and uncontrollably oxidized to the undesired $Sn^{4+}$ non-perovskite phase. A related process is observed when converting metallic Au into a $Cs_2(Au^{I}Au^{III})Cl_6$ perovskite using a (poly)chlorides that are formed in aqua-regia ($HNO_3$+HCl) together with CsCl salt.[39]

These examples suggest that polyhalides, or polyhalide salts, may play a dual role in both the degradation and the regeneration of HaPs. By understanding and controlling HaP formation through the interaction between polyhalides and $M^0$ metals, one can provide two significant added values to the field of halide-perovskites. First, the use of elemental metals in the synthesis should be industrially beneficial. As mentioned earlier, controlling the transformation of metals into functional semiconductors offers many advantages, especially since metals can be deposited in large-scale roll-to-roll processes, are easy to store, and implement.[29] Second, by controlling the metal-oxidation reaction, we can prevent or reverse the spontaneous occurrences of photolysis and electrolysis. This will allow us to engineer efficient operation, regeneration, and recycling strategies, ultimately enhancing the functional durability and operational lifetime of HaP-based technologies.

Hence, in this work, we aim to understand the conversion chemistry of elemental metals into metal-halide perovskites and perovskite-related bromides using a poly*bromide* melt (**PBM**). More specifically, this study investigates methylammonium-based polybromide melts, formulated as $MABr_{2n+1}$ species, and examines their reactivity with *four* different *M*-site metals: **Pb, Sn, Au,** and **Cu**. The work investigates the generality of the *M*+PBM conversion process, the underlying formation chemistry, its intrinsic limitations, and possible strategies to overcome them. We note that the decision to focus on polybromides rather than polyiodides, which was demonstrated on Pb-based systems,[31,32] was initially based on the novelty of polybromides for the conversion of metals to functional metal-halide perovskites, with clear potential applicability across different *A*-*M*-Br perovskite systems. Nevertheless, the successful outcome of this study goes far beyond this particular demonstration, establishing a new universal approach to perovskite synthesis and marking a paradigm shift that positions it as a cornerstone for future studies.

As will be emphasized below, the different metals differ substantially in their oxidation stability and accessible oxidation states. This is a critical limitation for direct metal-to-perovskite conversion, because although all four metals form known perovskite or perovskite-related halide structures, their thermodynamically preferred oxidation states do not always naturally match the charge requirements of the target HaP framework. In particular, successful conversion requires that the metal species ultimately adopt an effective 2+ average oxidation state compatible with the corresponding bromide perovskite or perovskite-related phase. We note that choosing Br-based perovskite systems was motivated by novelty and potential applicability,[1–4,40] but a similar approach can be taken for I- or Cl-based perovskite systems.

We show that, under common PBM-based reaction conditions, *all four* metals can be converted into the corresponding bromide perovskite or perovskite-related phases while reaching a required oxidation state. This demonstrates the potential universality of the PBM-mediated conversion approach, extending much beyond previous related reports[31,32,38], and highlights its potential importance for understanding and controlling HaP formation and stabilization. Overall, we believe that this study provides a foundation for developing new synthesis routes and stabilization strategies for HaP-based technologies.

## 2. Methods

Experimental details are provided in the SI, Appendix S.G. Since the manuscript focuses on the methylammonium polybromide melt as a central result, an exception is made for its synthesis, which is presented in Section 3.1. Essential characterization details are included in the main text in cases where it was considered to be essential for the understanding of the result.

## 3. Results

In this part of the manuscript, we present our main results, focusing on the synthesis and characterization of the polybromide melt (Section 2.1), and its reactivity with Pb, Sn, Au and Cu (Section 2.2). Section 2.2 is divided into three subsections, each using a different reaction approach: reaction with a thin metal film via spin-coating as an initial proof-of-concept (2.2.1), *in-situ* measurement of a reaction between a metal powder and the PBM in an isolated chamber, where we study the reaction propagation (2.2.2), and a stoichiometric-controlled reaction via grinding, where we study the role of the PBM composition (2.2.3). Finally, our results have been filed for a patent application number 202531085179 (published in 2025).

### 3.1. Synthesis of a methylammonium polybromide melt

The methylammonium polybromide melt (PBM) was prepared using methylammonium bromide salt (MABr), liquid bromine ($Br_2$), and hydrobromic acid (HBr – 48 wt% in $H_2O$) (see **Figure 1a**). Our starting PBM composition used an equivalent mole ratio of MABr to $Br_2$, which had to be heated to be liquified.

When adding roughly 10 mol% of HBr (overall, a 1:1:0.1 mole ratio between MABr:$Br_2$:HBr; or mixing 112 $mg$ of MABr, 52 $\mu l$ of $Br_2$ and 11.3 $\mu l$ of 48 wt% HBr in $H_2O$), the mixture liquified into a viscous, shiny, reddish-brown melt, forming our "baseline" PBM. We will see that in some cases, such as Sn, the baseline PBM composition is too aggressive, and since the redox chemical activity of the PBM effectively originates from the reduction chemical potential of the bromine (alone or part of the polybromide), altering the $Br_2$ content or the $Br_2$:HBr ratio, will show modified chemical activity. For example, for Sn we will use a "mild" PBM compound, containing a 1:0.25:1 mole ratio between MABr:$Br_2$:HBr, which will allow us to drastically reduce over-oxidation of $Sn^0$ into its undesired $Sn^{4+}$ state.

Without HBr, mixing MABr with $Br_2$ yields an orange solid, with reduced but evident $Br_2$ brown vapor (see **Figure 1a**). Upon heating, the compound melts while simultaneously undergoing substantial $Br_2$ gas release. The addition of HBr allows us to overcome two issues: first, it melts at room temperature, allowing us to react it with metal films via coating processes, such as spin-coating; second, it compounds with free bromine more effectively, essentially eliminating visual evidence of brown $Br_2$ vapor (unless heated). Therefore, the addition of a minimal amount of HBr (>0.07 mole ratio with respect to a 1:1 MABr:$Br_2$ mixture) provides a clear benefit in its use as a reagent. To avoid potential slow $Br_2$ vaporization of the PBM, we store the melt in a refrigerator (+4 °C) where it solidifies, allowing it to be stable for, at least, three months (the tested duration). More details regarding our preparation protocol for the PBM can be found in the SI, Appendix S.A.

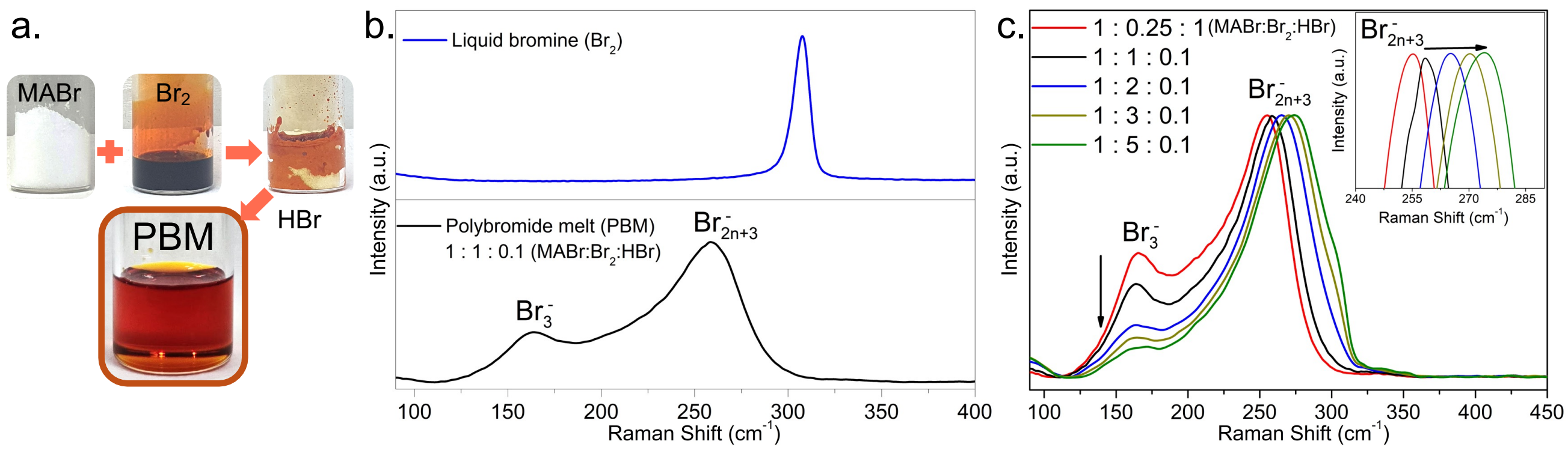

**Figure 1: Formation and characterization of the PBM.** (*a*) Images of the precursors, intermediate products (solid), and final reddish-brown PBM. (*b*) Raman spectrum of the default [1:1:0.1] PBM showing characteristic peaks of $Br_3^-$ (163 $cm^{-1}$) and $Br_5^-$ (217, 260 $cm^{-1}$), which are clearly distinct from the $Br_2$ peak at 307 $cm^{-1}$ (*cf.* SI, Appendix S.A, Table S.A1). (*c*) Comparison between Raman spectra of PBM with increasing $Br_2$:$Br^-$ ratio; *black arrows* indicate corresponding changes in $Br_3^-$ : $Br_{2n+3}^-$ peak intensity ratio and $Br_{2n+3}^-$ peak shift (*inset*).

To verify the presence of a polybromide in the PBM liquid phase, we performed Raman spectroscopy. **Figure 1b** shows strong, broad peaks at 163 $cm^{-1}$, 217 $cm^{-1}$ (shoulder) and 260 $cm^{-1}$, which are attributed to $Br_3^-$ and higher-order polybromide species such as $Br_5^-$, $Br_7^-$, $Br_9^-$, etc.[41–43], respectively. From the Raman result, we can confirm the validity of the following polybromide formation equation:

$$\text{Eq. 2} \quad Br_2 + Br^- \leftrightarrow Br_3^-$$

$$\text{Eq. 3} \quad nBr_2 + Br_3^- \leftrightarrow Br_{2n+3}^- \;\; (n \geq 1)$$

Eq. 2 represents the initial formation of the polybromide, while Eq. 3 represents the extension of the polybromide into a higher order that requires more than one $Br_2$ molecule per $Br^-$ anion.

We note that even in our default PBM, where $Br_2$:$Br^-$ is effectively <1, we observe Raman peaks that relate to higher-order $Br^-_{2n+3}$ polybromide. Since the presence of the 260 $cm^{-1}$ band indicates a high order of the polybromides, which require an excess of $Br_2$, one can assume that the $MA^+$ cation itself may be coordinated with the polybromides as $MA(Br_{2n+1})$, but that it should also stay in some portion coordinated with its native bromide as MABr dissolved in the PBM.

Interestingly, if we increase the $Br_2$:$Br^-$ ratio, Raman measurements show systematic changes between the $Br_3^-$ and $Br^-_{2n+3}$ peak ratio and $Br^-_{2n+3}$ peak position (see **Figure 1c)**. In addition to the Raman spectrum changes, the color of the PBM becomes darker as we increase the $Br_2$:$Br^-$ ratio, where, similar to the methylammonium polyiodide,[32] PBM also has a characteristic band-to-band-like transition (see SI, Appendix S.A , Figure S.A2). While referring to the 1:1:0.1 PBM as our default, we will see that a PBM with a higher $Br_2$:$Br^-$ ratio leads to a stronger oxidation activity of the PBM, and can serve as an important tuning factor in controlling the transformation of the metals into their corresponding metal halide perovskites using PBM. Raman and optical transmission spectra can then provide indirect measures to quantify PBM quality and activity and allow one to follow desired (or undesired) modifications in the PBM and its potential activity.

### 3.2. Converting Metals to HaP with PBM:

In this section, we explore the reaction between Pb, Sn, Cu and Au, and the PBM. We start by characterizing the resulting films after spin-coating PBM onto PVD-deposited metallic films, focusing on identifying the product's phases, oxidation state, and morphology. We will complement these results with a more systematic study of the chemistry, using controlled ratios between the PBM and the metals, using metallic powders, which will allow us to trace intermediate phases and evaluate the response to changes in the PBM composition. Since we are essentially activating a corrosion reaction, with the PBM acting as the oxidizing agent, we will pay attention to the redox chemistry between the metals and the PBM. Later, we show that redox potentials can serve as a useful predictive metric for Metal–PBM reactivity, targeting the correct oxidation state (2+ or alternating 1+/3+) for forming a HaP; this will be particularly important for the case of Sn.

#### *3.2.1. PBM-assisted metal-to-HaP conversion via spin-coating*

To examine the reaction between PBM and the different $M^0$ metals (Pb, Sn, Au, Cu), we spin-coated the PBM onto the metallic films, with PBM deposition occurring as the metallic film spun. The spin-coating process was performed in an ambient atmosphere and is illustrated and photographed in **Figure 2a** and Figure S.B1, respectively. The metallic films themselves were PVD-deposited to a thickness of a few hundred nm – see details in SI, Appendix S.G.

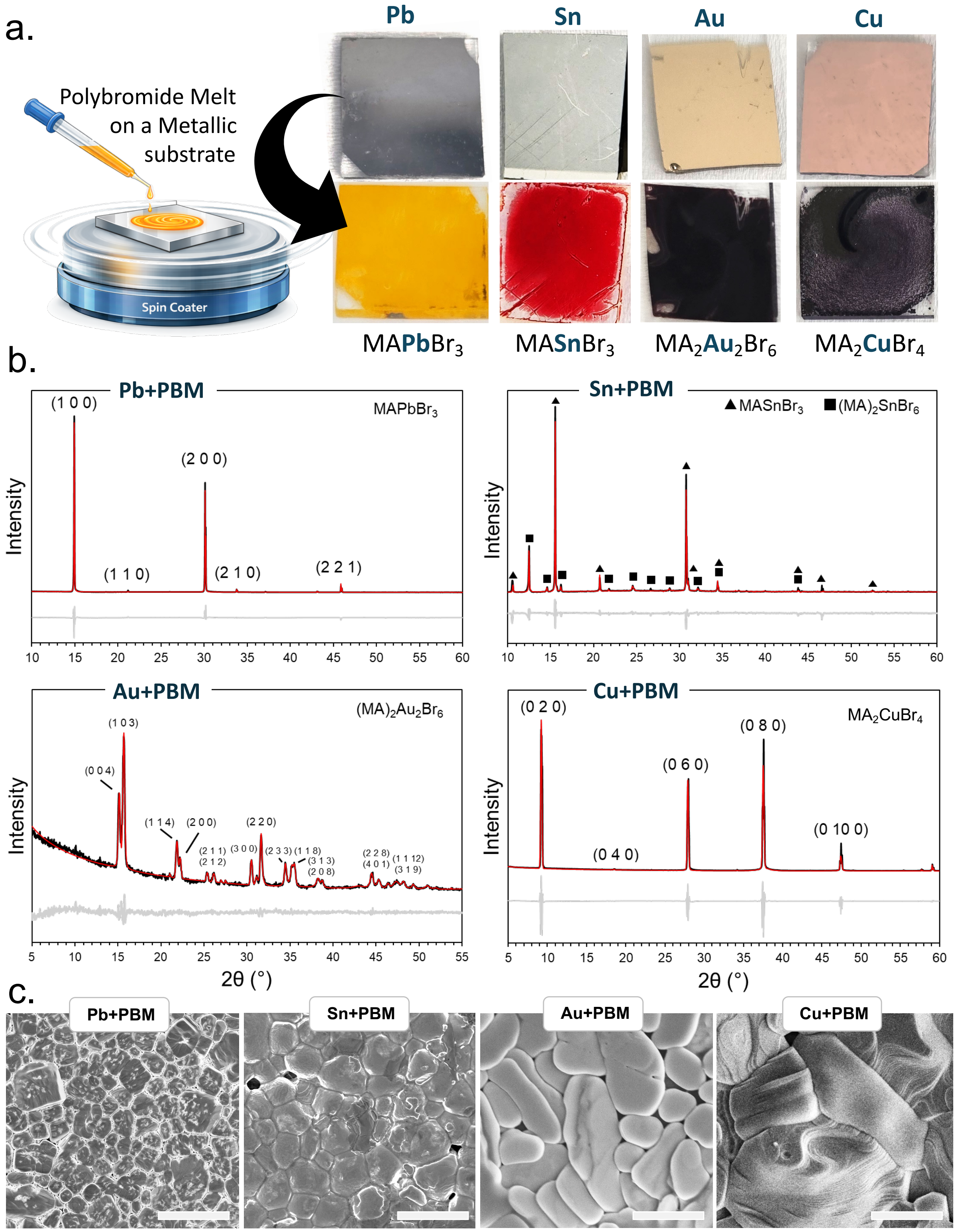

**Figure 2: Thin-film reaction with PBM via spin-coating.** (*a*) Illustration of the PBM-assisted metal-to-HaP conversion via spin-coating and actual photographs of PVD-deposited metallic films before and after the conversion reaction. The actual spin-coating process is shown in Figure S.B1. (*b*) Powder XRD from the converted films, and their Rietveld refinement. Further details on the refinement results, including goodness-of-fit parameter and the March–Dollase preferred-orientation factor, are reported in the SI, Appendix S.B. (*c*) Characteristic SEM micrographs from the converted films presented in *a*; the scale bar in all the micrographs corresponds to 5 $\mu m$.

Regardless of the chosen metal, we ended up forming films that changed color to closely resemble MA-*M*-Br-based perovskites, namely $MAPbBr_3$, $MASnBr_3$, $MA_2Au_2Br_6$, and $MA_2CuBr_4$ for *M*=Pb, Sn, Au, and Cu, respectively (see **Figure 2a**). When analyzing the corresponding optical transmission and photoluminescence spectra (see SI, Appendix S.B), we found a good agreement with previously reported optical responses for these MA-*M*-Br perovskites. To identify the different phases unambiguously, we analyzed XRD data that was collected from the converted films (see **Figure 2b** and further refinement details in the SI, Appendix S.B). Following Rietveld refinement, which included a March–Dollase correction to account for preferred orientation, we verified that *all* films yielded the desired perovskite phase with no or minimal residual metal (no metal was observed on the back side of the coated substrate).

As for the morphology, we found it varies significantly between the different metals (see **Figure 2c**). The $MA_2Au_2Br_6$ films show relatively large crystallites, but with less sharp facets. The $MA_2CuBr_4$ films exhibit the largest crystallites with a clear layered structure, as expected for a 2D perovskite system and as previously observed following the reaction between $CuBr_2$ and MABr[46]. $MAPbBr_3$ and $MASnBr_3$ produce densely packed films, with a typical morphology to spin-coated films.[44,45]

While being rather continuous, we still found pinholes in the Pb- and Sn-originated film and a significant amount of the undesired $MA_2SnBr_6$ phase. Guided by the powder-grinding results (see Section 3.2.3), using a mild PBM, with a 1:0.25:1 ratio between MABr:$Br_2$:HBr, **Figure 3** shows that Sn-originated films can result in tightly packed grains, similar to those found in high-quality devices, and almost complete elimination of the $Sn^{4+}$ $MA_2SnBr_6$ phase. We should note that this process was done entirely outside of a glove-box, suggesting a potential route for perovskite synthesis in an ambient environment. XPS analysis (Figure S.B11) also revealed a significant decrease in the $Sn^{4+}$: $Sn^{2+}$ ratio at the film surface when both films were exposed to air for a similar amount of time. A caveat for this optimization is that unreacted Sn remained, which may be an issue for some applications. Nevertheless, one can think of it as a potential Schottky contact between the source metal and the resulting semiconductor.

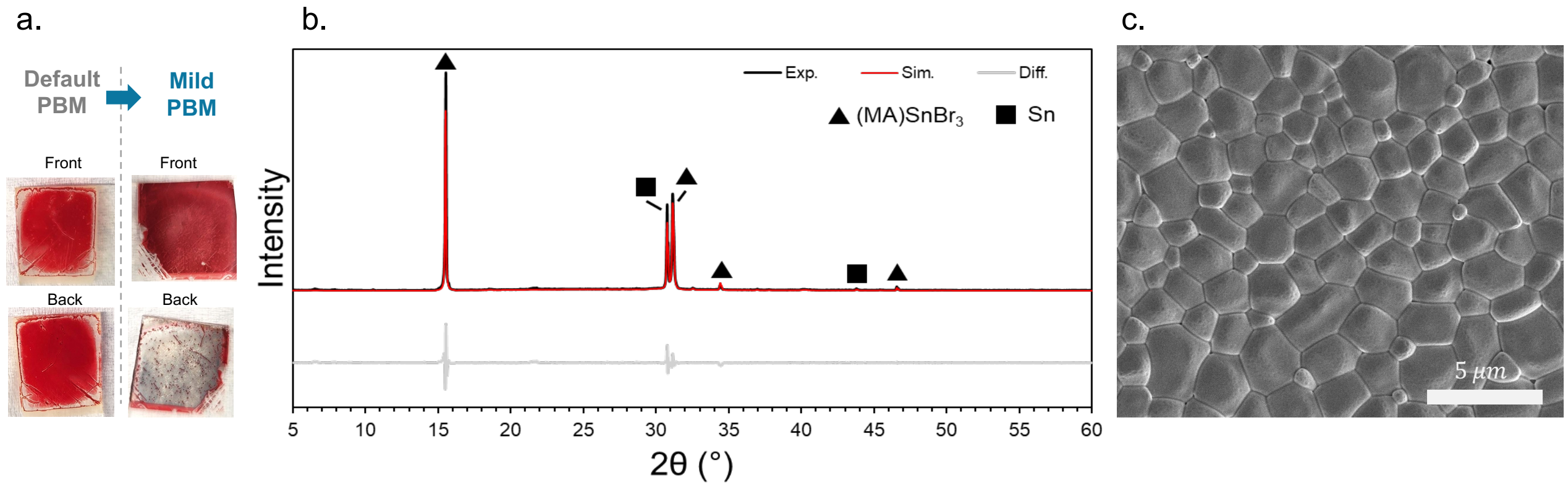

**Figure 3**: **Sn metal film conversion with the mild PBM via spin-coating:** (*a*) Visual comparison between the reacted Sn film from the front and back of the glass onto which the Sn film was evaporated. The reactions took place via spin-coating of the default (1:1:0.1) and mild (1:0.25:1) PBM compounds. The mole ratios refer to MABr:$Br_2$:HBr. (*b*) Powder XRD from the converted films; for Rietveld refinement analysis, see SI, Appendix S.B, Figure S.B10. (*c*) Characteristic SEM micrographs from the converted Sn film, showing a much better morphology with respect to the reaction with the baseline PBM (*cf.* **Figure 2b,c**).

Overall, we have verified that a single reaction protocol, in which metal films serve as precursors, can generate a series of HaPs with the same generic method – exposure to a 1:1:0.1 PBM. While further optimization is required to achieve functionally optimal films by tailoring the film morphology, PBM composition, and/or reaction timing, this demonstration is of utmost importance as it reveals a possible pathway towards industrialization and demonstrates the potential to become a generic approach for synthesizing a variety of HaPs from metal films that are today deposited industrially in roll-to-roll fashion.

#### *3.2.2. Following the reaction evolution: in-situ reactivity between the metallic powder and PBM*

To gain insight into the mechanism of the metal conversion reaction, we performed a set of sequential XRD measurements on the metal powders immediately after exposure to the PBM solution. Into the sealed reaction chamber, which was constructed from two X-ray transparent Mylar® films clamped between two concentric Teflon™ coils, we added a fine metal powder and a drop of PBM, which were then immediately sealed and measured in a consecutive series of timed measurements. In **Figure 4** we can see the photographed reaction chambers after the measurements, and a series of XRD patterns that show the phase evolution from the metal (bottom) to the final product (top). Unlike the thin-film reaction, which was open to air, the reaction chamber is sealed, so no oxygen- or humidity-related reactions are expected, except for the native water present in HBr or residual oxygen trapped between two Mylar sheets. We note that the *in-situ* reaction conditions did not allow us to extract reaction rates, but rather a qualitative evaluation of the sequence of phase emergence and transition.

Based on the XRD patterns in **Figure 4**, we make *three important observations*. First, within the experiment time resolution (~15 min), in the cases of Pb, Sn and Cu, we saw the metallic peak slowly disappearing and the perovskite phase appearing, whereas in the Au case it happened immediately. Second, similarly to the thin-film reaction, in all cases, the final product we observed was a halide perovskite phase, namely: $MAPbBr_3$, $MASnBr_3$, $MA_2Au_2Br_6$, and $MA_2CuBr_4$ for *M*=Pb, Sn, Au, and Cu, respectively.

Finally, in the cases of Sn and Cu, we observed intermediate phases alongside the perovskite phases. For Cu, the transient $MA_3Cu_2Br_5$ phase (see Rietveld refinement in Figure S.C2) forms and then later disappears, leaving only the perovskite $MA_2CuBr_4$ phase. For Sn, we confirmed that the transient phase is $MA_2SnBr_6$ (see Rietveld refinement in Figure S.C3**)**. Interestingly, the $MA_2SnBr_6$ phase, which is a phase with the more oxidized $Sn^{4+}$ cation, disappears, leaving behind the perovskite $MASnBr_3$ phase, suggesting a comproportionation reaction between metallic $Sn^0$ and $Sn^{4+}$, which forms a $Sn^{2+}$ phase.

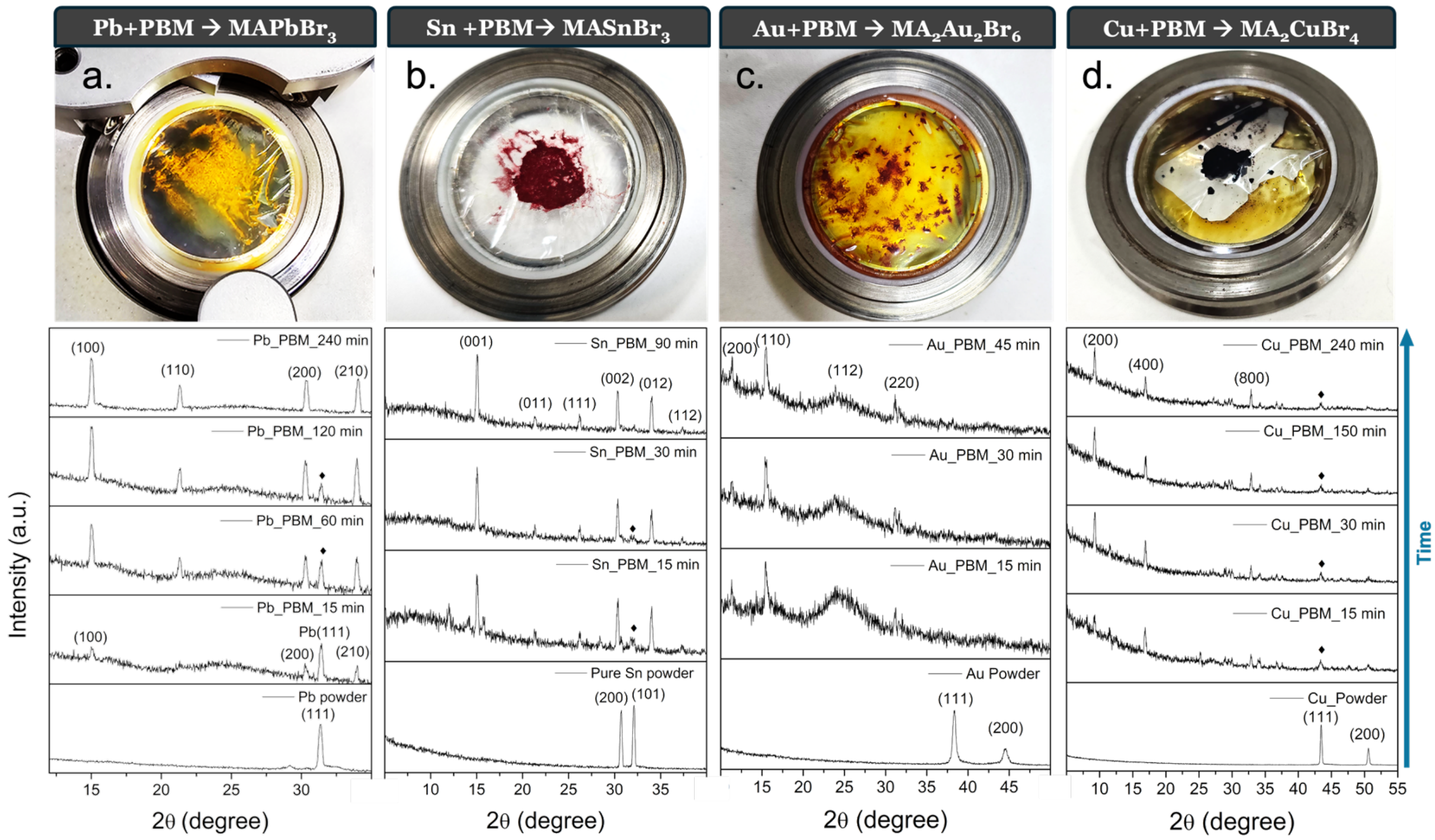

**Figure 4: *In-situ* reaction experiments.** Stepwise *in-situ* XRD measurements that follow the conversion reaction of (*a*) Pb, (*b*) Sn, (*c*) Au, and (*d*) Cu metal powder with PBM between two Mylar foils. The bottom diffractogram in each column refers to the pure metal encapsulated between Mylar foils. After adding the PBM drop, we immediately sealed the entire system (*M*+PBM) with a top Mylar foil, then began performing sequential XRD measurements. The time noted in each box is the duration from the moment the drop was added. Each measurement iteration took ~15 min. The background diffractions from the Mylar and Mylar+PBM are shown in the SI, Appendix S.C, Figure S.C1.

Unexpectedly, when the metallic reactant is exhausted, and the comproportionation reaction can no longer happen, the $Sn^{2+}$ perovskite phase remains. Given the oxidative nature of the PBM, it is likely that the $MASnBr_3$ phase will propagate further to the more oxidized Sn phase, namely $MA_2SnBr_6$. To further investigate this, we isolated this reaction by growing $MASnBr_3$ crystals (following ref. [47]), and reacting them with PBM within the same *in-situ* reaction chamber (no Sn metal included). As shown in Figure S.C4 (SI, Appendix S.C), we confirmed that the perovskite phase further oxidizes to the $MA_2SnBr_6$ phase. While not all the $Sn^{2+}$-perovskite oxidized to a $Sn^{4+}$, the stabilization factors between the PBM and the perovskite phase remain unresolved and require a more systematic study, which will follow this one.

### 3.2.3. *Optimizing reactivity: reacting ground metal powder with PBM*

To study the chemistry between the metals and the PBM more carefully, we performed more controlled reactions by mixing and grinding fine metal powders (Pb, Cu, Au, and Sn) with PBM mixtures, keeping the metal: MABr mole ratio at 1:1. Since HBr already contains water, we performed the reaction in air. Each metal powder was thoroughly ground with varying PBM compositions using a mortar and pestle until a visually homogeneous mixture was obtained. The resulting products were subsequently analyzed by powder XRD, and the diffraction patterns were refined to quantify the phase composition (see results in **Figure 5** and the refinement results in SI, Appendix S.D).

We should note the case of Sn, which was an exception among the other metals. In the case of Sn, the reaction with the default [1:1:0.1] PBM was violently exothermic, with visible bubbling and the release of orange $Br_2$ vapor. Also, the product was almost exclusively white, characteristic to $MA_2SnBr_6$ (similar to the periphery of the Sn film after spin-coating shown in **Figure 2a**). To reduce the reactivity, we found that diluting the PBM with additional HBr (specifically from mole ratio of 0.1 to 1.0) led to a much milder reaction and the formation of the dark-colored $MASnBr_3$ phase. Overall, we essentially reduced the $Br_2$ activity by adding HBr to form a milder PBM reactant, where the additional HBr likely acted as a diluent for the reactive PBM.

As shown in **Figure 5**, the reaction with the initial baseline PBM compositions (1:1:0.1 for Pb, Cu, and Au, and 1:1:1 for Sn) was not optimal, as in every case we found unreacted MABr, unreacted metals, metal halides, or other competing phases. Sn under the milder PBM condition was a notable exception yielding over 90 % of the desired $MASnBr_3$ perovskite phase, all the other metals yielded less than 50 % of the perovskite phase, with Au-perovskite yielding slightly above 15%.

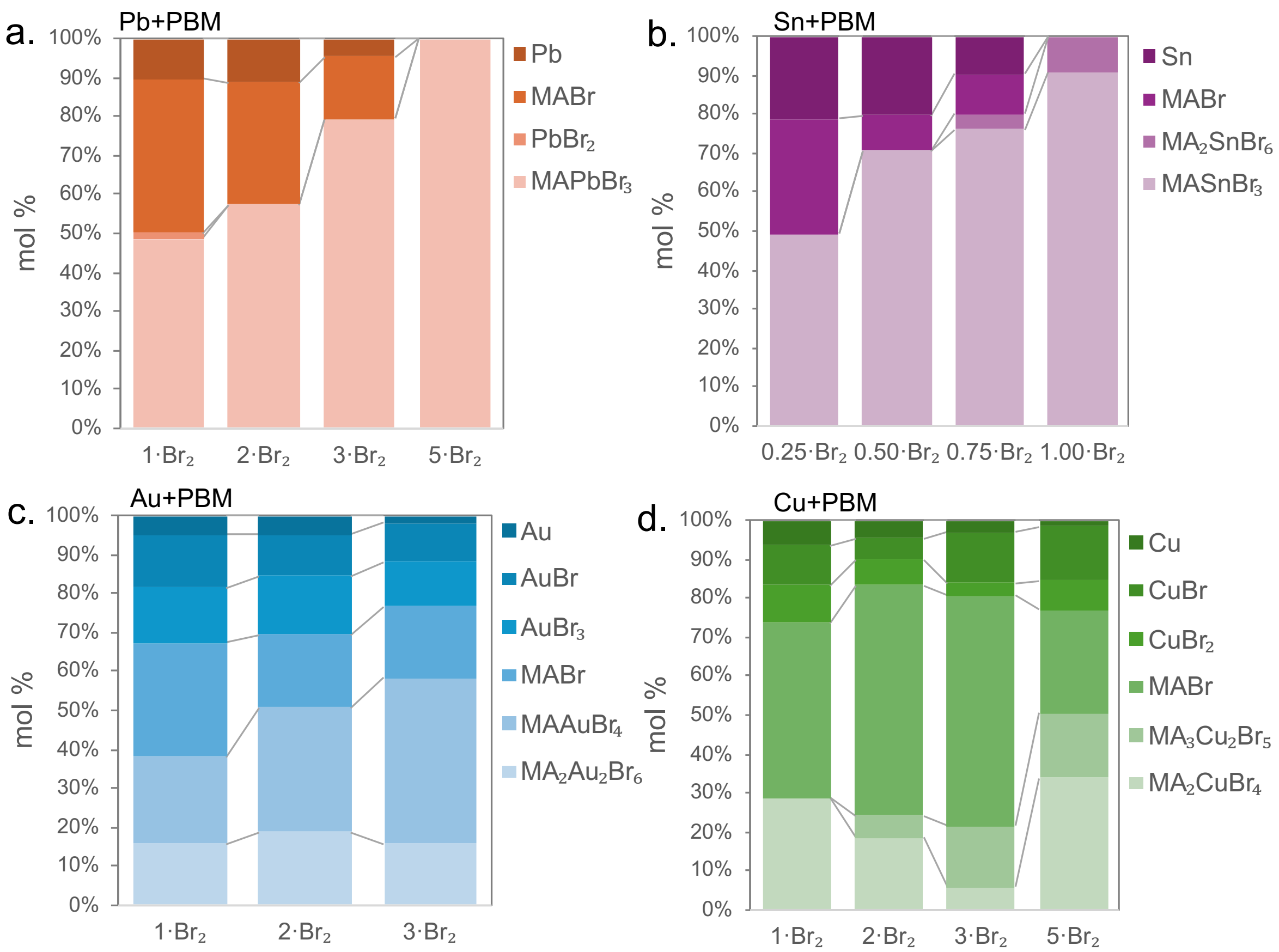

**Figure 5**: **Metal powder + PBM grinding experiments**. Rietveld refinement-based phase fraction analysis after reactions between (*a*) Pb, (*b*) Sn, (*c*) Au, and (*d*) Cu metal powders and the PBMs with varying $Br_2$ contents. For Pb, Cu and Au we kept the MABr:HBr ratio constant at 1:0.1 while increasing the $Br_2$ mole ratio from 1 to 2, 3, or 5. For Sn, we kept the MABr:HBr ratio constant at 1:1, while decreasing the $Br_2$ mole ratio from 1 to 0.75, 0.50, and 0.25. In all cases, the [metal:MABr] mole ratio was 1:1. The numerical tabulated results are found in SI, Appendix S.D, Table S.D1, and the Rietveld refinements are shown in Figure S.D1 and Figure S.D2.

Besides declaring this method as generally inefficient for producing HaPs, we used this setup to investigate the role of $Br_2$ content on the formed products. If our goal was to maximize the yield of the perovskite phase, increasing the $Br_2$ content should have increased the reactivity of Pb, Cu, and Au. For the case of Sn, since we also gained ~10 % of the undesired $MA_2SnBr_6$ phase, we decided to further *decrease* the content of $Br_2$ as a way to decrease the reactivity, and thus potentially increase the fraction of the desired $MASnBr_3$ phase. For the rest of the metals, we kept the MABr:HBr ratio constant at 1:0.1 while further increasing the $Br_2$ mole ratio from 1 to 2, 3, and 5. After gaining powder XRD data and refining it (see SI, Appendix S.D), we calculated the phase ratios and related trends shown in **Figure 5**.

As shown in **Figure 5**, all cases exhibit a systematic decrease in the content of the precursor metals as the $Br_2$ content increase. As for the gradual improvement in the yield of the desired perovskite phase, a systematic trend was observed only for Pb and Sn (increasing $Br_2$ content led to higher yields of the $MAPbBr_3$ and $MASnBr_3$ phases). Only when using Pb did this grinding method result in a pure perovskite phase. Interestingly, for Sn, we found that while the total amount of the perovskite phases decreases as the $Br_2$ content decreases, the undesired $MA_2SnBr_6$ phase also disappears, leaving $MASnBr_3$ as the only perovskite phase present.

Au and Cu show a far more complex phase assemblage than Pb and Sn. In the case of reaction with Au, the phase assemblage includes AuBr, $AuBr_3$, and $MAAuBr_4$ alongside the $MA_2Au_2Br_6$ perovskite. $MAAuBr_4$, which was found with the largest portion among the product phases, is of particular interest among the competing Au phases. We note that the $MAAuBr_4$ structure was derived from the $MAAuCl_4$ analog[48]; for details, see SI, Appendix S.E.

For Cu, the phase assemblage included CuBr, $CuBr_2$, and $MA_3Cu_2Br_5$ alongside the perovskite phase $MA_2CuBr_4$. The only clear trends that were observed when increasing the $Br_2$ content were the decrease of the Cu metal precursor, and an increase in the $Cu^{1+}$ -related to the formation of the $MA_3Cu_2Br_5$ phase. $Br_2$ is an oxidizing agent, so to explain the inconsistent behavior a more reasonable explanation would be degradation. Following Lee et al.,[40] $MA_2CuBr_4$ would undergo a reversible amorphization in response to humidity. With increased humidity, the formation of an amorphous phase naturally reduces the observed phase fraction in XRD measurements, leading us to mistakenly believe that the $MA_2CuBr_4$ content decreases relative to the other phases. In addition, they showed that humidity will react with $MA_3Cu_2Br_5$ to form $CuBr_2$ ($MA_3Cu_2Br_5 \rightarrow MA\uparrow + CuBr_2$). This can explain the sudden increase in the content of $MA_2CuBr_4$ at 5·$Br_2$ reaction: degradation of $MA_3Cu_2Br_5$ to $CuBr_2$ and reaction of $CuBr_2$ with MABr that leads to $MA_2CuBr_4$. This interpretation is consistent with the observed disappearance of MABr, $CuBr_2$, and $MA_3Cu_2Br_5$ alongside the appearance/increase of $MA_2CuBr_4$. Overall, counterintuitively, we see that humidity may promote the formation of $MA_2CuBr_4$, but also that the correct way to control these reactions is under strictly humidity-controlled conditions.

Overall, we understand that grinding is a useful tool for exploring reactivity trends, but not very useful for a production route. We note that MABr was always present (except for the $5{\cdot}Br_2$ ratios when reacting Sn and Pb), which could be explained as either an incomplete reaction or a spontaneous degradation. While we do not have a complete explanation for all observed phases, one possible source of these findings is that the grinding was not done in an air-tight environment, nor under controlled humidity conditions. Hence, if $Br_2$, due to its volatile nature, has escaped during the grinding process, precursors such as unreacted metal and MABr would remain but would also disappear as we add more $Br_2$, as observed. We can also conclude that the thin-film (**Figure 2**) or *in situ* (**Figure 4**) experiments, which were faster (spin-coated) or air-tight (*in situ*), did not show comparable degradation of binaries (metal-halides) and precursors.

## 4. Discussion

The results of this study establish methylammonium poly*bromide* melt (PBM) as a reactive medium capable of converting a variety of elemental metals (Pb, Sn, Au, Cu) into bromide perovskite or perovskite-related phases under a common chemical framework. Considering previous works[31,38], in which methylammonium poly*iodide* melt was found to convert metallic Pb to form Pb-HaPs phases, the expansion into the bromide-polyhalide and the demonstration on a variety of metals suggests that polyhalides play a significant role in the chemistry, and possibly the stabilization, of halide-perovskite systems.

A central outcome of this work is the surprising generality of the PBM-mediated conversion route. Despite major differences in redox behavior and oxidation-state stability, Pb, Sn, Au, and Cu all yielded bromide perovskite or perovskite-related products upon reaction with PBM. This is nontrivial because oxidation-state control remains a principal chemical constraint in halide perovskite synthesis, particularly for systems such as Sn and Au, where their ideal perovskite phase oxidation states are not necessarily stable under ambient conditions. The ability of a common reactive medium to promote conversion across these chemically distinct metals indicates that PBM functions not only as an oxidizer but also as a stabilization factor in the formation of the metal-bromide perovskite frameworks.

Nevertheless, the results show that while the overall chemical concept is general, the details of the specific reaction matter. Comparison of the reaction configurations and conditions between the thin-film, *in situ* powder, and ground-powder experiments isolates distinct aspects of the chemistry and clarifies the nature of the PBM-mediated metal conversion. The thin-film geometry establishes the synthetic relevance of the approach for HaP film fabrication; *in situ* measurements identify plausible phase-evolution pathways and reveal intermediate phases; the controlled powder reactions define the PBM compositions that favor or suppress specific products. Considered together, they show that PBM-mediated metal-to-perovskite conversion is governed by the type of metal, the content of the PBM reagent, and the environmental conditions of the reaction. We should note that most of the reactions (except the *in situ*) were performed in open air, which likely led to oxygen- or humidity-mediated phases, such as $MA_3Cu_2Br_5$ and $MA_2SnBr_6$.[40,49] And yet, the spin-coating-assisted reaction on the thin films appears to yield good phase purity, despite being performed in ambient conditions. Clearly, a complementary study of the reactivity under inert conditions is warranted and will be conducted in future work.

Notably, changes in the PBM's content allowed us to tune the kinetics and the resulting products, making PBM a tunable reactive medium rather than a single-purpose reagent. Within this framework, successful perovskite formation requires control over the extent of oxidation and an understanding of which products are stabilized under the chosen chemical conditions. Optical transition and Raman spectroscopy, where for the latter the $Br_3^-$:$Br_{2n+3}^-$ peak ratio and $Br_{2n+3}^-$ peak position (see **Figure 1c**), can be used as measures for the $Br_2$:$Br^-$ ratio, and figures of merit to the oxidation activity of a PBM composition. The Sn system illustrates this balance most clearly. Under the standard PBM composition, the reaction produced both $MASnBr_3$ and the more oxidized $MA_2SnBr_6$ phase, whereas lowering the $Br_2$ activity and increasing the HBr content shifted the reaction toward substantially improved $MASnBr_3$ phase purity (see **Figure 5**). The *in situ* measurements are consistent with the conclusion that Sn-perovskite formation requires carefully balanced redox conditions: sufficient oxidation to convert metallic Sn, but limited over-oxidation to avoid formation of $MA_2SnBr_6$.

Overall, the results from the thin-film reaction with the low-$Br_2$-content PBM phase (**Figure 3**), where almost pure $MASnBr_3$ phase was observed together with Sn metal, strongly suggest that a comproportionation reaction, meaning reducing $Sn^{4+}$ to $Sn^{2+}$ by oxidizing $Sn^0$ to $Sn^{2+}$, plays a critical role in stabilizing the $Sn^{2+}$ phase. Similar to the grinding experiment, lower PBM reactivity left behind a clear layer of un-reacted Sn film, which can also be explained by a comproportionation reaction and explains the increase in $Sn^{2+}$: $Sn^{4+}$ phase ratio. Overall, these insights can lead to new directions for stabilization of $Sn^{2+}$ phases at ambient conditions.

### 4.1. Metal–PBM reaction guidelines

Using four metals with distinct electrochemical activities allows us to establish general thermodynamic guidelines for predicting metal–PBM reactivity and the resulting oxidation states. In this section, we identify the electrochemical trends that rationalize our observations and provide a framework for future metal-polyhalide chemistry.

Let us consider a general redox reaction between a metal *M* and a polybromide melt that forms an $AMX_3$ HaP:

$$\text{Eq. 4} \quad M + AX_3 \rightarrow AMX_3$$

This overarching process can be broken down into two half-cell redox reactions, where the polybromide provides the oxidative driving force:

$$\text{Eq. 5} \quad M \rightarrow M^{2+} + 2e^-$$

$$\text{Eq. 6} \quad X_3^- + 2e^- \rightarrow 3X^-$$

Following the redox exchange, phase stabilization occurs. The newly formed Br− ligands coordinate with the oxidized metal to form intermediate complexes (e.g., $[MX_3]^-$),[50] which subsequently react with the $A^+$ cation to precipitate the corner-sharing $[MX_6]$ octahedral network:

$$\text{Eq. 7} \quad M^{2+} + 3X^- \rightarrow [MX_3]^-$$

$$\text{Eq. 8} \quad [MX_3]^- + A^+ \rightarrow AMX_3$$

While Eqs. 7 and 8 dictate the ultimate structural stability of the perovskite phase, the initial and most critical bottleneck for successful conversion is the thermodynamic feasibility of the metal oxidation.

Let us review the expected oxidation states for the selected metals. Pb strongly favors the $Pb^{2+}$ state. Au typically favors $Au^{3+}$, as $Au^+$ tends to disproportionate into $Au^0$ and $Au^{3+}$.[27] Sn can form $Sn^{2+}$, but the $Sn^{4+}$ state is thermodynamically favored and often forms spontaneously under ambient conditions. [49,51] Cu is relatively stable in both 1+ and 2+ states.

To rationalize the observed reactivity, we calculate the electrochemical cell potentials and corresponding Gibbs free energies by coupling the reduction of $Br^{3-}$ with the oxidation of the selected metals (**Table 1**). We note that the oxidation fundamentally occurs with $Br_2$, where $Br^-$ is a complexing agent to $Br_2$ that does not have much effect on the overall oxidation potential, where in both cases its reduction potential is reported around 1.07-1.09V [28],[52]. Similarly, the redox potential of $Br_5^-$ is similar to that of $Br_3^-$.[41] Since the analysis in **Table 1** is rather crude, we will refer to the potential of the $Br_3^-$ found in reference [52].

**Table 1:** Electrochemical potentials for halogen / polyhalides reduction and Au, Sn, Cu, and Pb metal oxidation with respect to SHE. 'a' and 'b' correspond to ref. [28] and [52], respectively, where 'cal.' represents that the potentials were calculated from the literature values. The Gibbs free energies are calculated from $\Delta G/F = -nE$ where $E$ is the electrochemical potential and $n$ is the number of electrons.

| | Reaction type | Reaction Equation | half-cell potential | Cell potential (for Ox- types: reaction with $Br_3^-$) | Energy per electron | Reaction electrons | Reaction free energy | Final stable product | Desired oxidation |
|---|---|---|---|---|---|---|---|---|---|
| | | | [V] | [V] | [eV] | [#] | [eV] | | |
| Br | Red | $Br_3^- + 2e^- \leftrightarrow 3Br^-$ | 1.08 b | | | | | | |
| Pb | Ox-1 | $Pb \leftrightarrow Pb^{2+} + 2e^-$ | 0.13 a | 1.21 | -1.21 | 2 | -2.42 | + | + |
| | Ox-2 | $Pb \leftrightarrow Pb^{4+} + 4e^-$ | -0.78 cal. | 0.30 | -0.30 | 4 | -1.20 | | |
| | Ox-12 | $Pb^{2+} \leftrightarrow Pb^{4+} + 2e^-$ | -1.69 a | -0.61 | 0.61 | 2 | 1.22 | | |
| | disp. | $2Pb^{2+} \leftrightarrow Pb + Pb^{4+}$ | | -0.91 cal. | | 4 | 3.64 | | |
| Sn | Ox-1 | $Sn \leftrightarrow Sn^{2+} + 2e^-$ | 0.14 a | 1.22 | -1.22 | 2 | -2.44 | | + |
| | Ox-2 | $Sn \leftrightarrow Sn^{4+} + 4e^-$ | 0.00 cal. | 1.08 | -1.08 | 4 | -4.30 | + | |
| | **Ox-12** | $Sn^{2+} \leftrightarrow Sn^{4+} + 2\,e^-$ | -0.14 [a] | 0.93 | -0.93 | 2 | -1.86 | | |
| | **disp.** | $2Sn^{2+} \leftrightarrow Sn + Sn^{4+}$ | | -0.15 [cal.] | | 4 | 0.58 | | |
| Au | **Ox-1** | $Au \leftrightarrow Au^{1+} + e^-$ | -1.69 [a] | -0.61 | 0.61 | 2 | 1.22 | | + |
| | **Ox-2** | $Au \leftrightarrow Au^{3+} + 3\,e^-$ | -1.50 [cal.] | -0.42 | 0.42 | 6 | 2.50 | | + |
| | **Ox-comp-1** | $Au + 2Br^- \leftrightarrow AuBr_2^- + e^-$ | -0.96 [a] | 0.12 | -0.12 | 2 | -0.24 | | + |
| | **Ox-comp-2** | $Au + 4Br^- \leftrightarrow AuBr_4^- + 3e^-$ | -0.85 [a] | 0.23 | -0.23 | 6 | -1.36 | + | + |
| | **Ox-12** | $Au^{1+} \leftrightarrow Au^{3+} + 2\,e^-$ | -1.40 [a] | -0.32 | 0.32 | 2 | 0.64 | | |
| | **Ox-comp-12** | $AuBr_2^- \leftrightarrow AuBr_4^- + 2\,e^-$ | 0.80 [cal.] | 1.88 | -1.88 | 2 | -3.76 | | |
| | **disp.** | $3Au^{1+} \leftrightarrow 2Au + Au^{3+}$ | | 0.19 [cal.] | | 3 | -0.58 | | |
| | **disp. w/ comp** | $3AuBr_2^- \leftrightarrow 2\,Au + AuBr_4^- + 2Br^-$ | | 0.11 [cal.] | | 3 | -0.32 | | |
| Cu | **Ox-1** | $Cu \leftrightarrow Cu^{1+} + e^-$ | -0.52 [a] | 0.56 | -0.56 | 2 | -1.12 | | |
| | **Ox-2** | $Cu \leftrightarrow Cu^{2+} + 2e^-$ | -0.34 [cal.] | 0.75 | -0.75 | 2 | -1.49 | + | + |
| | **Ox-12** | $Cu^{1+} \leftrightarrow Cu^{2+} + e^-$ | -0.15 [a] | 0.93 | -0.93 | 2 | -1.86 | | |
| | **disp.** | $2Cu^{1+} \leftrightarrow Cu + Cu^{2+}$ | | 0.19 [cal.] | | 2 | -0.37 | | |

Based strictly on the sum of the half-cell potentials, the $Br^{3-}$ polybromide possesses sufficient oxidative power to formally drive Sn to $Sn^{4+}$, Cu to $Cu^{2+}$, and Pb to $Pb^{2+}$. It does not, however, possess the potential to spontaneously oxidize Au on its own. For Au oxidation to yield a negative free energy, the complexation of Au with bromine ligands (forming $[AuBr_2]^-$ or $[AuBr_4]^-$) must be factored into the thermodynamic driving force, ultimately favoring the $Au^{3+}$ state. We note that, if a metal spontaneously compounds with the halide ligand, the free energy should be pushed down even further, regardless of the metal. Therefore, as with Au, adding *M--Br* compounds will make the free energies more negative; however, the chemical trends, and with them the predictive power, will not change, and are therefore omitted.

While the PBM naturally drives Cu and Pb to the exact oxidation states required for their respective perovskite system, it theoretically possesses the oxidative power to over-oxidize both Sn and Au to states ($Sn^{4+}$ and pure $Au^{3+}$) that do not accommodate standard 3D or 2D perovskite structures. Yet, our reaction protocol successfully yielded the required oxidation states for perovskite formation across all four metals. This apparent thermodynamic contradiction can be explained by a redox-balancing mechanism in the case of Sn, most notably *comproportionation*, or a kinetically-stable phase extraction via precipitation in the case of Au.

Let us consider Sn first. The total free energy heavily favors the oxidation of Sn to $Sn^{4+}$ (forming $MA_2SnBr_6$). However, if unreacted, the positive free energy for the disproportionation reaction says that a spontaneous comproportionation reaction occurs between the metal and the over-oxidized product, meaning $Sn^0 + Sn^{4+} \leftrightarrow 2Sn^{2+}$. This dynamic explains the contrasting results between the thin-film reactions using the baseline PBM (**Figure 2**) and the mild PBM (**Figure 5**). The mild PBM lacked the oxidative capacity to fully consume the underlying Sn film. The remaining metallic Sn at the interface actively hindered the propagation of the $Sn^{4+}$ state by continuously driving the comproportionation reaction, thereby kinetically trapping and stabilizing the desired $Sn^{2+}$ ($MASnBr_3$) perovskite phase. Conversely, in the *in situ* XRD experiment (**Figure 4**), we likely see the comproportionation buffering in action, where the $Sn^{4+}$ phase is present but then disappears. Once no metallic Sn was present (Figure S.C4**)**, the system lost this protective comproportionation buffer, allowing the PBM to drive the system toward the more oxidized $Sn^{4+}$ state.

Unlike Sn, the case of Au comproportionation cannot explain the phase stability since the disproportionation free energy is negative. Since complexation with the halides is what thermodynamically allows $Au^0$ to convert to $Au^{3+}$, a rapid precipitation of the rigid perovskite lattice, kinetically trapping the mixed $Au^{1+}/Au^{3+}$ valence state within the $MA_2Au_2Br_6$ framework, is one possible explanation. Alternatively, the $MA_2Au_2Br_6$ system may be thermodynamically stabilized around (poly)halides, but here, electrochemical reasoning will have to make room for ligand-field theories and accurate *ab initio* calculations.

Overall, while this electrochemical framework does not account for the exact lattice formation energies of the final crystalline compounds, the half-cell potentials serve as highly effective, chemically intuitive guidelines. They accurately predict the necessary oxidative thresholds and highlight how tuning the halogen activity within the PBM can precisely control the final oxidation state of the metal, ensuring successful halide perovskite conversion.

As a final step, after obtaining our most optimized chemical composition of the PBM from the different experiments, we measured Raman spectroscopy. The results (see SI, Appendix S.F, Figure S.F1) indeed support that the obtained phases correspond to previously reported Raman spectra from the relevant perovskite phases. It is also notable from the comparison between the spectra that $MAPbBr_3$ and $MASnBr_3$ have a significant central peak due to their anharmonic nature[53]. In contrast to these, $MA_2CuBr_4$ and $MA_2Au_2Br_6$ phases have no central Raman peak, as expected from classical rigid semiconductors. This result demonstrates that our findings can support fast progress in the fundamental understanding of a large variety of HaPs through comparative studies, alongside the clear technological merit it holds as a potential method to break through inhibitors for industrial development, such as organic solvent toxicity and minimizing waste.

## 5. Conclusion

In summary, we demonstrate that a methylammonium polybromide melt (PBM) can serve as a general reactive medium for converting elemental metals into bromide perovskite and perovskite-related phases. Using a common PBM-based framework, elemental Pb, Sn, Cu, and Au were transformed into $MAPbBr_3$, $MASnBr_3$, $MA_2CuBr_4$, and $MA_2Au_2Br_6$, respectively, across thin-film and powder-based reaction configurations. This establishes that metal-to-perovskite conversion is not limited to a specific metal or halide chemistry but can instead emerge from a broader polyhalide-driven redox and phase-stabilization pathway.

Beyond demonstrating synthetic access, this work shows that PBM is a tunable reaction medium. By varying its composition, the oxidation strength and reaction outcome can be shifted substantially, allowing improved control over phase purity and product selectivity. This was especially evident in the Sn system, where suppressing $Br_2$ activity and increasing HBr content favored $MASnBr_3$ over the more oxidized $MA_2SnBr_6$ phase. The combined thin-film, *in situ*, and ground-powder experiments further reveal that successful conversion is governed by an interplay between metal redox chemistry, reagent composition, and phase stabilization under the chosen reaction conditions.

A key conceptual outcome of this study is that metal oxidation potentials, together with the possibility of disproportionation or comproportionation, provide a useful first-order framework for predicting whether PBM-assisted conversion will yield the oxidation states required for perovskite formation. Although this electrochemical picture does not yet account explicitly for all compound-specific thermodynamic and kinetic contributions, it captures the major trends observed experimentally and offers chemically intuitive design rules for extending this chemistry to additional metal-halide systems.

Overall, our findings are highly significant from both synthetic and mechanistic perspectives. Synthetically, this work demonstrates that diverse elemental metals (and not only those favorably forming a 2+ state, such as Pb) can serve as direct precursors for the formation of functional HaPs. Because these elemental metals are inherently stable and fully compatible with industrial, large-scale physical vapor deposition processes, this method bypasses the severe vulnerabilities and storage complexities associated with traditional metal-halide precursors. Mechanistically, illuminating the fundamental redox interactions between polyhalides and metals provides vital insights into the (photo)electrochemical degradation and regenerative self-healing mechanisms of operating devices. Whether the polyhalides are added intentionally or spontaneously formed through photolysis or electrolysis during device operation, they will mediate both the degradation and regeneration chemistry in halide perovskite systems through dynamic redox transformations involving metals and polyhalide salts.

Ultimately, the robust conversion pathways establish a new paradigm for perovskite synthesis. It opens realistic avenues for integrating high-quality perovskite films into advanced technological architectures, whether deployed as interlayers in next-generation tandem solar cells or light-emitting diodes, radiation detectors, humidity sensors, or even as active switching components in modular phase-change memristors for adaptive memory hardware that is based on electrochemically-formed conducting filaments.

## Acknowledgement:

Y.R. acknowledges the Ministry of Energy, Israel, Grant # 222-11-077, for their support in this project. Y.R. would also like to thank Prof. Davide Raffaele Ceratti from the Institut de Recherche de Chimie Paris for consultation regarding Au perovskite systems. A.M and Y.R. also thank Dr. Roxana Golan, Dr. Natalya Froumin, Vital Gelberg and Sonal Rajput, Amit Raz, Leen Gharra, Layla Khateeb, and Alon Gamshi for technical experimental support.

### Conflict of Interest:

The authors declare no financial or commercial conflicts of interest.

Supporting Information

# A Universal Polybromide Melt Strategy for the Direct Conversion of Metals into Optoelectronic-Grade Bromide Perovskites

Arindam Mondal[1], Mark Baranov[2], Lonia R. Friedlander[2], Mariela J. Pavan[2], Kirill Sobolev[1], Yevgeny Rakita[1,2]*

[1] Department of Materials Engineering, Ben Gurion University of the Negev, Be'er Sheva, Israel
[2] Ilse Katz Institute for Nanoscale Science and Technology, Ben-Gurion University of the Negev, Be'er Sheva, Israel

*Corresponding Authors – rakita@bgu.ac.il, arindamm@post.bgu.ac.il

## Table of Contents

# Appendix S.A: Polybromide melt (PBM): Supplementary Materials

## Suggested robust preparation protocol for a 1:1:0.1 PBM

Our optimized preparation protocol for forming a robust PBM reagent includes: (1) mixing MABr with 10 mol% of HBr in a glass vial with a Teflon seal; (2) slow addition of ice-cooled $Br_2$ to reduce $Br_2$ vapor pressure; (3) hand-shake mixing and a short sonication, after which the the visible $Br_2$ vapor quickly disappeared, resulting in a uniform reddish-brown solution. Heating the PBM led to the release of additional brown $Br_2$ vapor. Hence, to avoid potential slow $Br_2$ vaporization, we store the melt in a refrigerator (+4 °C), where it solidified. Before use, the solidified PBM was allowed to warm to room temperature until it re-melted..

## Raman

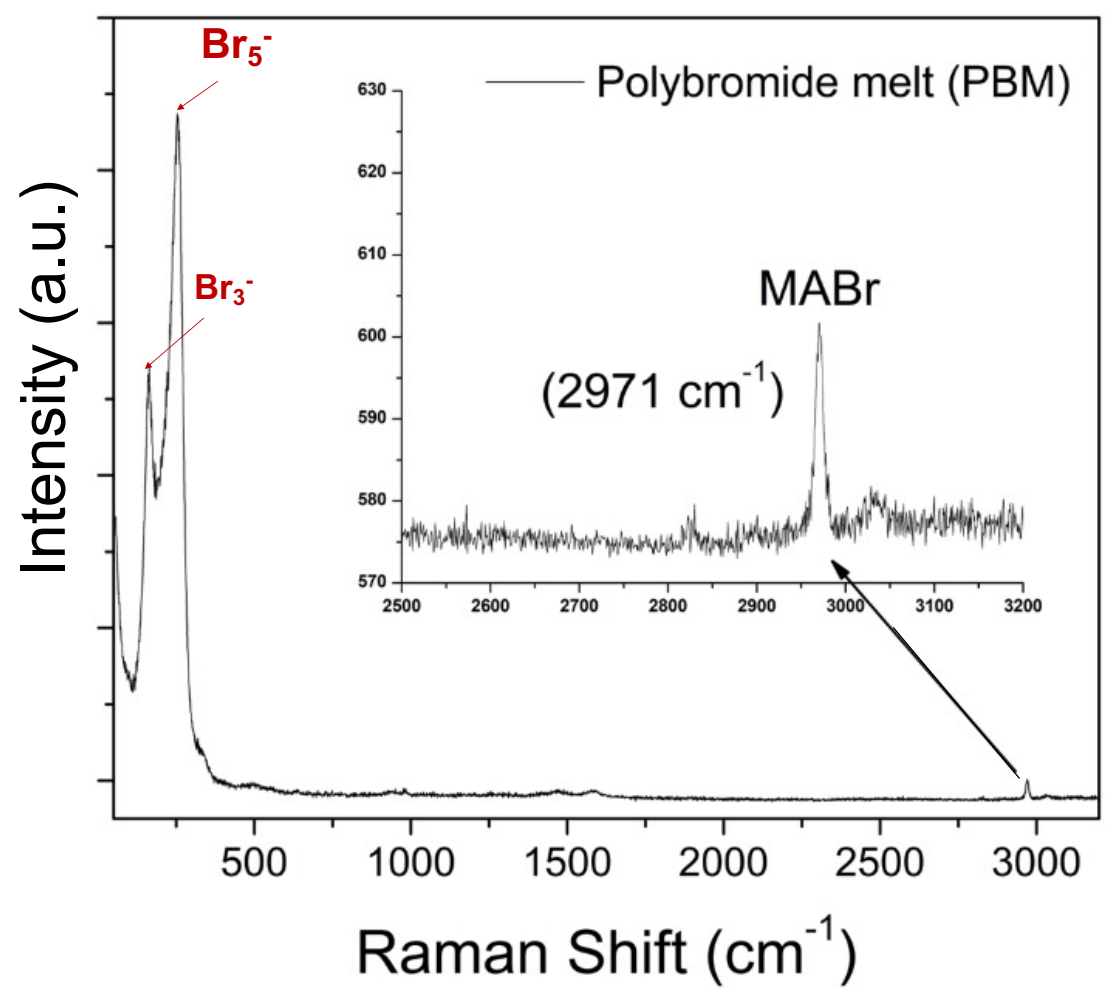

**Figure S.A1**: Full Raman spectrum of the PBM showing characteristic broad peaks at 162 cm$^{-1}$ for $Br_3^-$, and 218 (shoulder) and 257 cm$^{-1}$ for $Br_5^-$. An additional set of peaks between 2800-3100 cm$^{-1}$ (see inset) is associated with the $CH_3NH_3^+$ (MA+) cation.[1]

**Table S.A1**: Reported characteristic peaks for bromine and polybromides.

| Molecules | Reference Raman Shifts [cm$^{-1}$] | Associated bands | References |
|---|---|---|---|
| **$Br_2$** | 310 | Fundamental transition | [2] |
| **$Br_3^-$** | 163 | $\sum_g^+$ Symmetric Stretching | [3] |
| **$Br_5^-$** | 253 | $A_1$ Symmetric Stretching; | |
| | 210 | $B_1$ Asymmetric Stretching | [3] |

## Optical Transmission of PBM

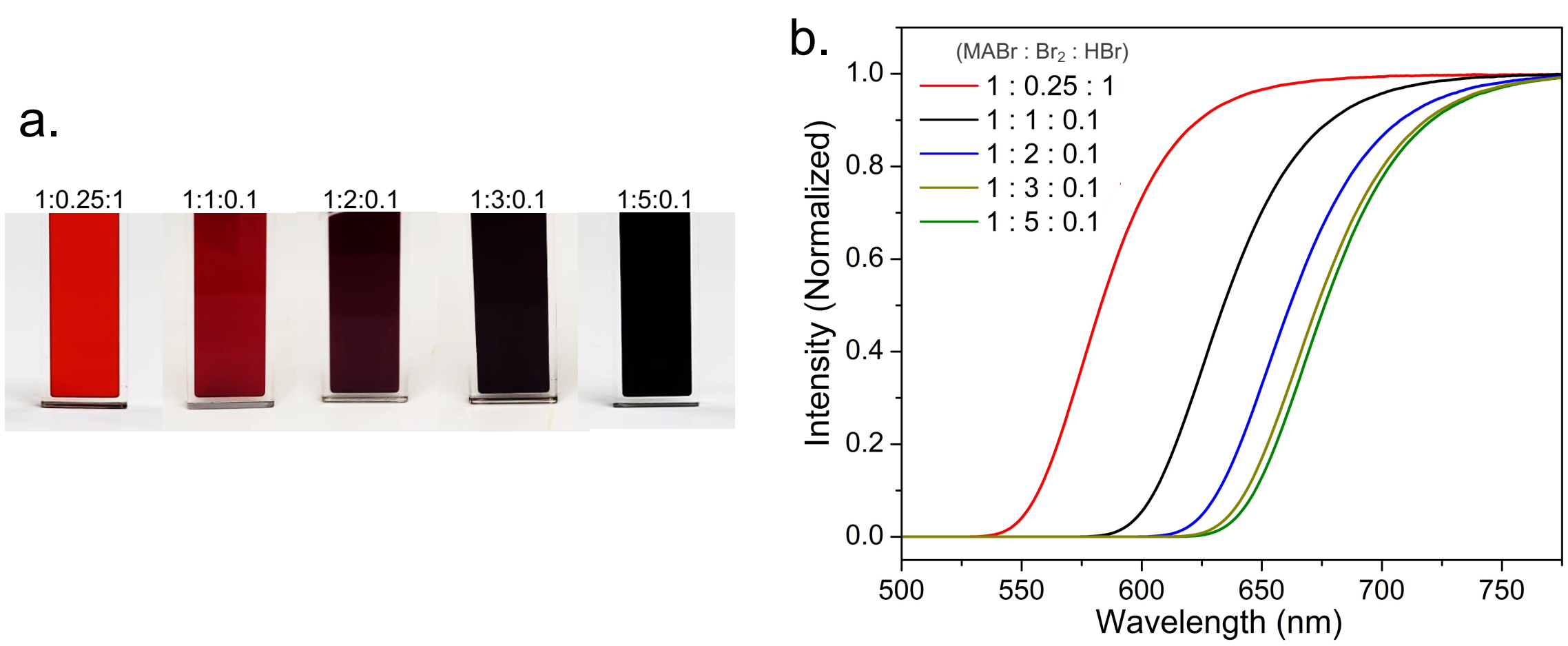

**Figure S.A2**: Optical transmission spectra from PBMs that contain different MABr:$Br_2$:HBr content (see legend), showing a band-to-band-like transmission and a systematic red-shift with increasing $Br_2$ content. The optical length of the solution in the Cuvettes was 1 mm for all five samples.

# Appendix S.B: PBM Spin-coated Metal-Film reaction: Supplementary Material

## Spin-coat experiment demonstration

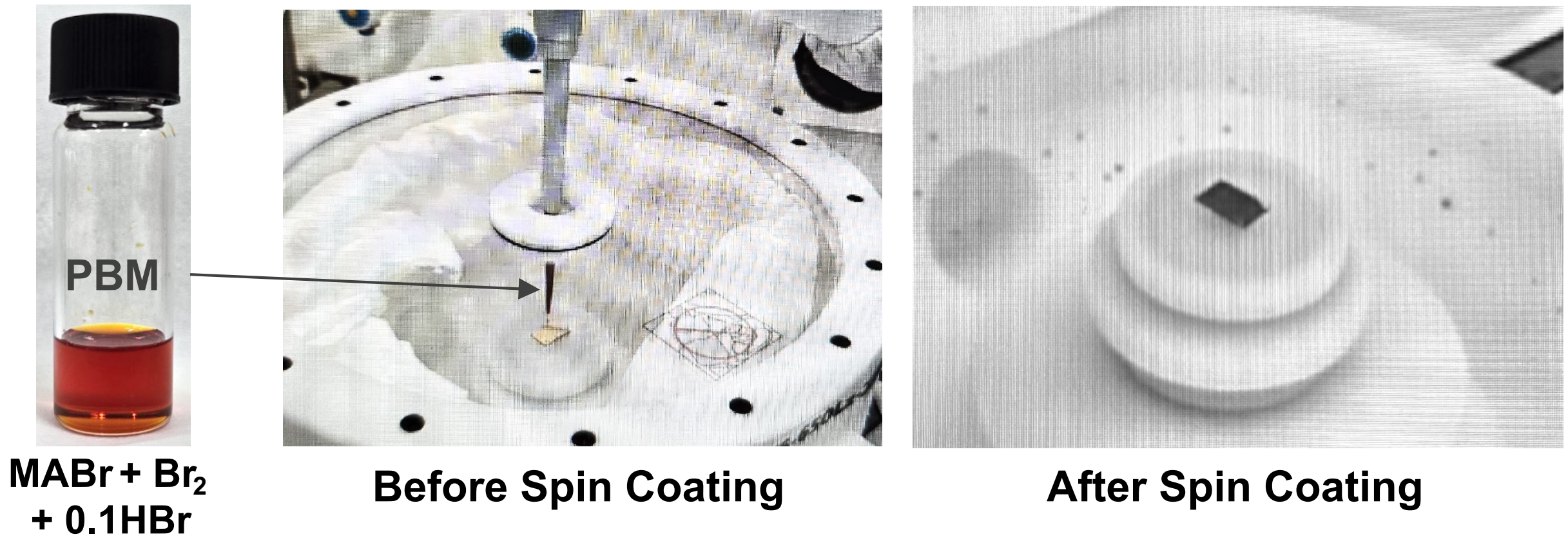

**Figure S.B1**: PBM implementation on an Au film via spin-coating, showing an instantaneous conversion of Au film (shiny yellow) into a dark, strongly absorbing film.

## Optical characterization

Here we present the optical characterization, specifically optical transmission and photoluminescence (PL), for the products of the *M*+PBM reactions. Tauc analysis is performed to determine the optical band gaps of the reaction product semiconductors[4].

**<u>Pb+PBM</u>**: We find an excellent agreement with $MAPbBr_3$ for the Transmission spectrum, bandgap, and PL.

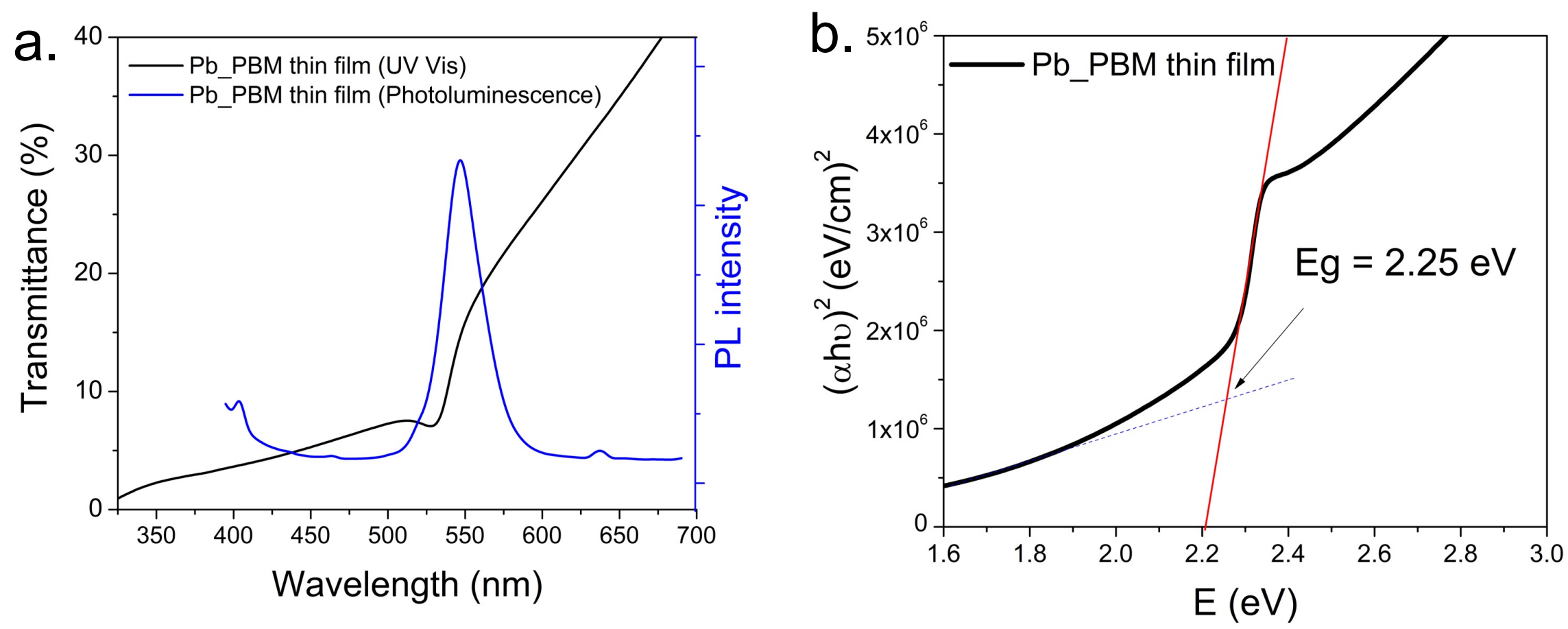

**Figure S.B2**: (a) optical transmission with an overlay of the PL spectrum (blue) from a product of a PBM-over-Pb film spin-coating reaction. (b) Tauc plot extracted from (a) to extract the bandgap value.

**Sn+PBM**: We find an excellent agreement with previously reported $MASnBr_3$ in both transmission spectrum and bandgap[5]. We did not detect significant PL, but this is expected in $MASnBr_3$ due to expected trap states, especially in the presence of $Sn^{4+}$ [5,6].

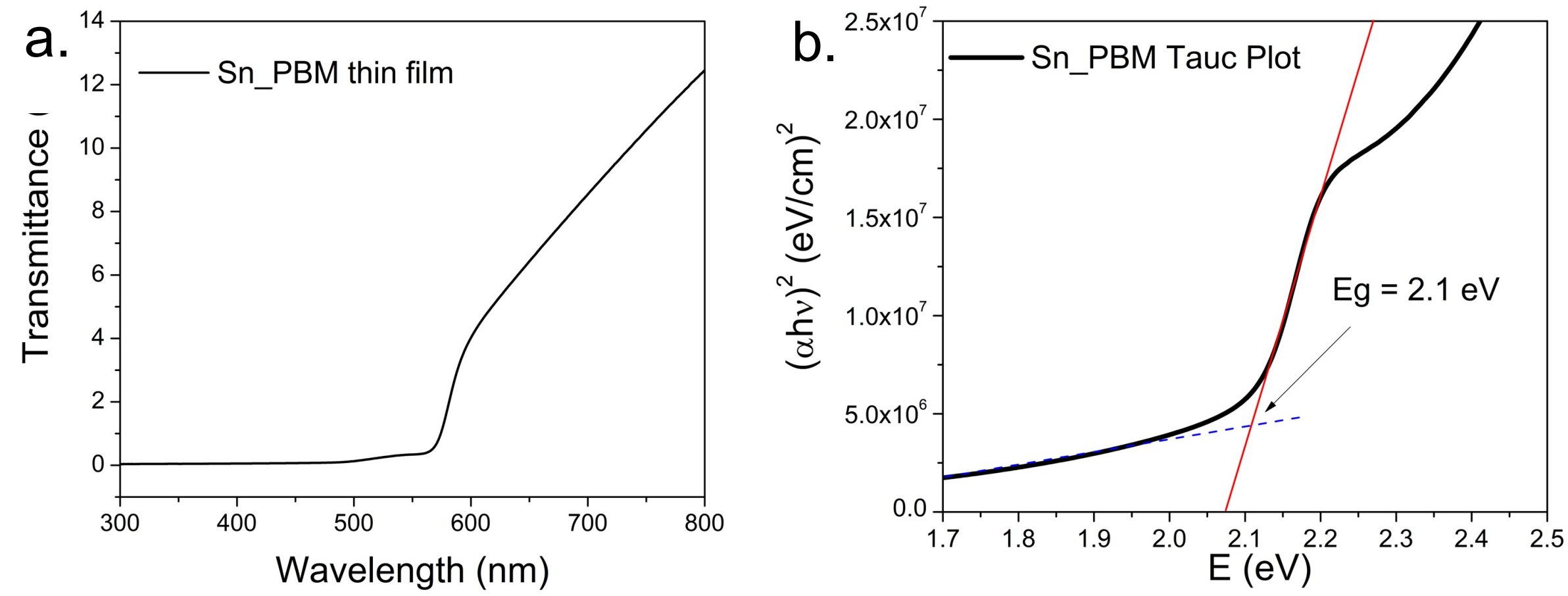

**Figure S.B3**: (a) Optical transmission from spin-coated reacted Sn with PBM. (b) Tauc plot extracted from (a) to extract the bandgap value.

**Au+PBM**: we found the transmission spectrum to agree with $MA_2Au_2Br_6$[7]. We note that unlike $Cs_2Au_2Br_6$ that does show measurable PL,[8] $MA_2Au_2Br_6$ does not show a significant PL intensity,[7] which is rationalized due to phononic disorder stabilization[9].

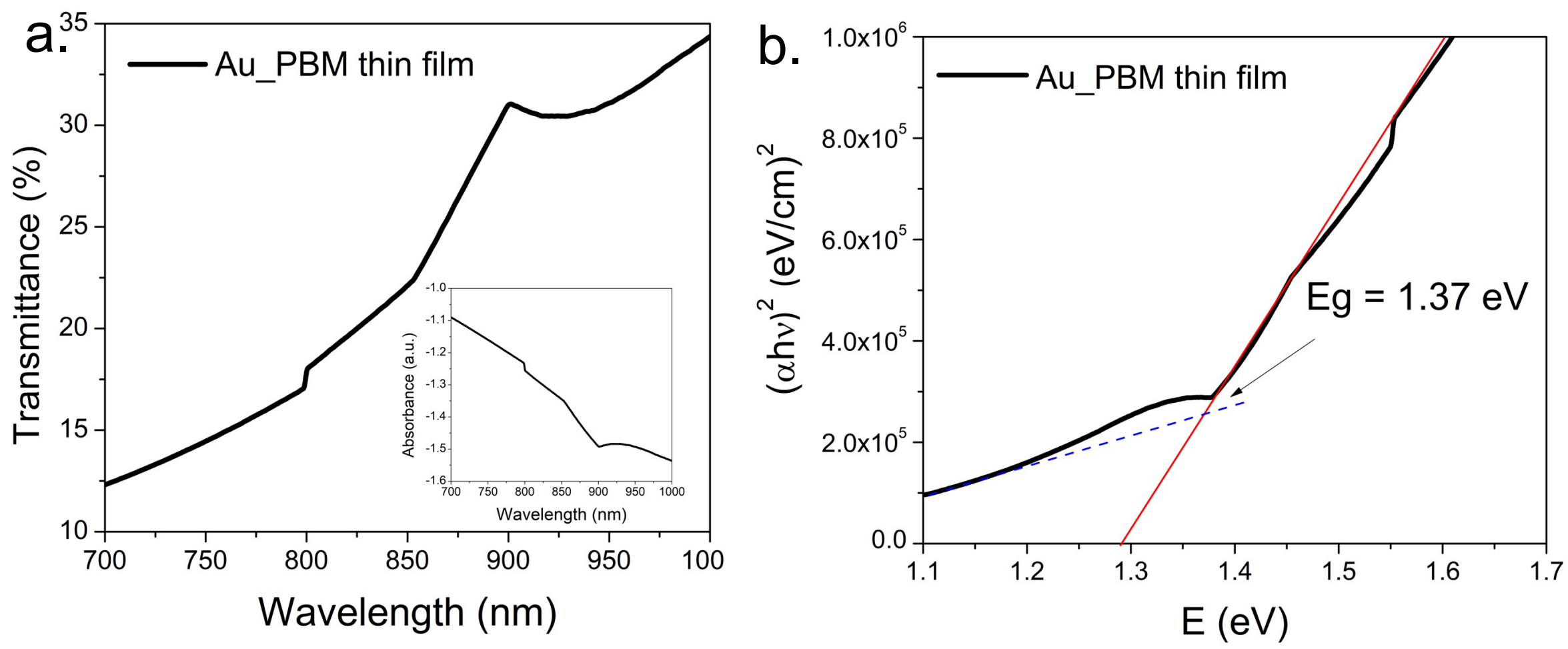

**Figure S.B4**: (a) optical transmission from a product of a PBM-over-Au film spin-coating reaction. (b) Tauc plot extracted from (a) to extract the bandgap value. The inset in (a) is the absorbance spectrum for comparison with reported absorbance plots[7].

**<u>Cu+PBM</u>**: The optical transitions of $MA_2CuBr_4$ are not a typical single band-to-band transition, but have additional Cu-Cu and Cu-Br molecular-like transitions.[10] Therefore, if we consider the complete absorption line shape, we find an excellent agreement with the previously reported absorption line shape of $MA_2CuBr_4$.[11] In addition, we find that its PL is also in good agreement with $MA_2CuBr_4$ [10,12].

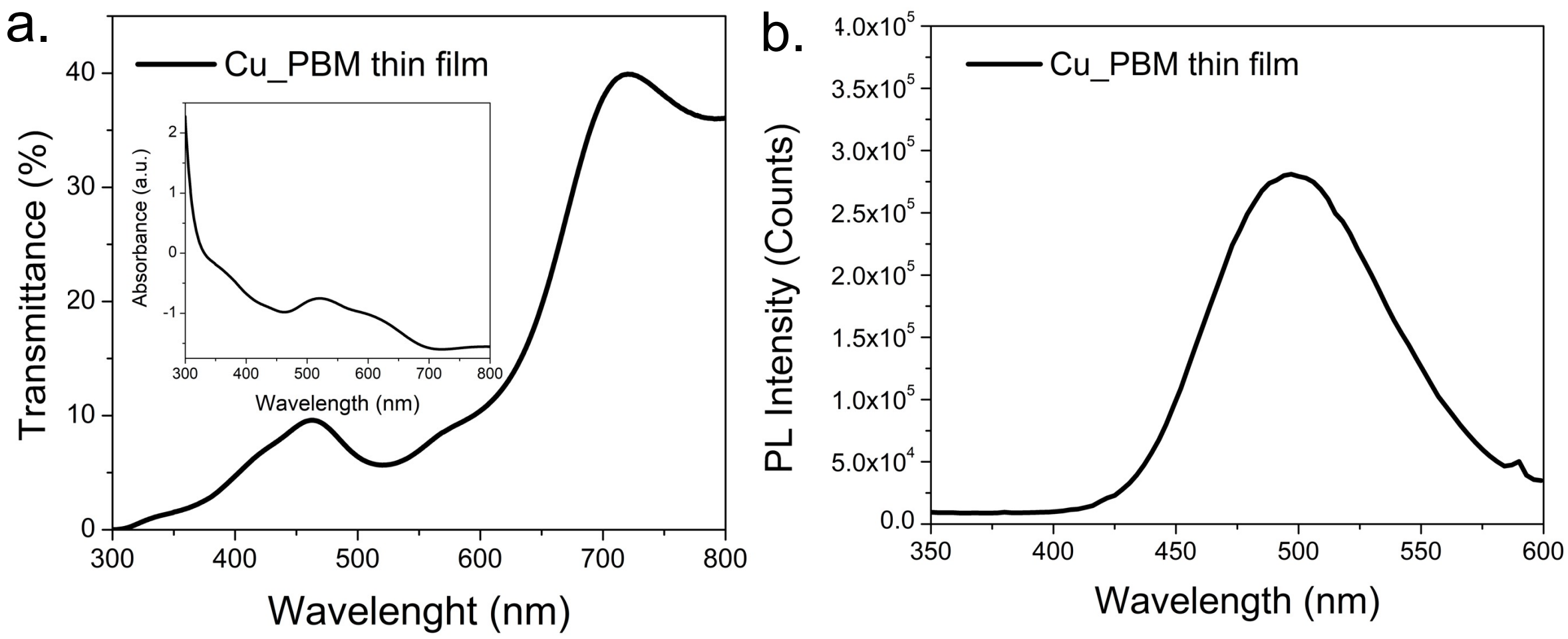

**Figure S.B5**: (a) optical transmission and (b) a PL spectrum from a product of a PBM-over-Cu film spin-coating reaction. The red inset in (a) is the absorbance spectrum that was added for easier comparison with reported absorbance plots[10].

## Powder XRD Rietveld refinement details for PBM-reacted metal films

**<u>Au</u>:**

$MA_2Au_2Br_6$ powder diffraction pattern (Figure S.B6) is consistent with the formation of the targeted methylammonium gold bromide phase, $MA_2Au_2Br_6$. The calculated profile reproduces the experimental peak positions and relative intensities well across the measured range, and the difference trace remains low and structureless, indicating good agreement between the structural model and the measured diffraction data.

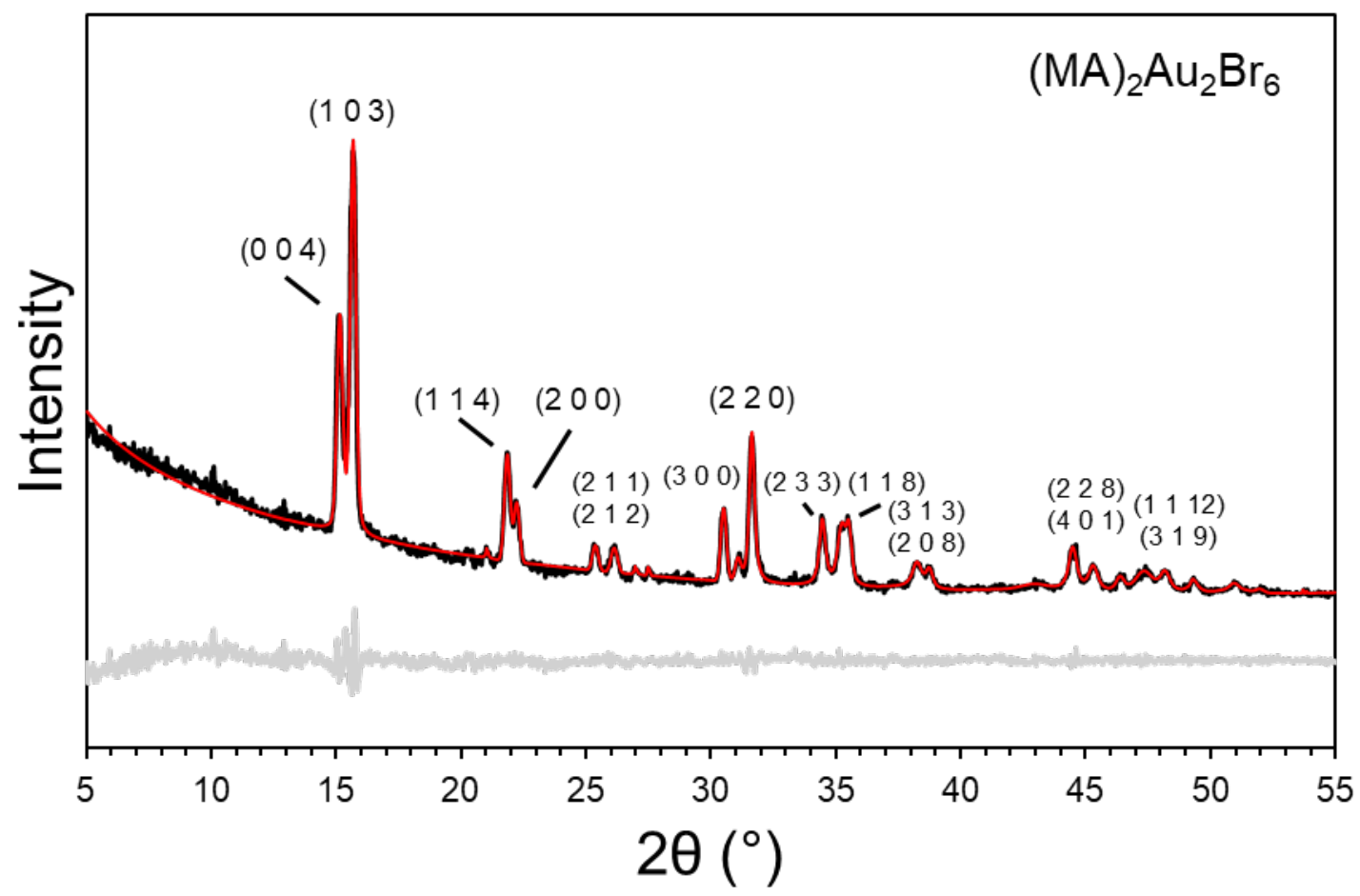

**Figure S.B6:** $MA_2Au_2Br_6$ (Rietveld refinement). Experimental (black) and calculated (red) powder XRD patterns for $MA_2Au_2Br_6$ with the difference curve shown in gray, confirming phase formation and good agreement between measured and simulated profiles. The Rietveld refinement converged to $R_{wp}$ = 8.49% and $\chi^2$ = 1.34.

**Cu:**

$MA_2CuBr_4$, diffraction pattern (Figure S.B7) supports the formation of methylammonium copper bromide as the dominant crystalline phase. Clear preferred orientation is evident and is accounted for using a March–Dollase correction (MD = 0.21 along [010]), again consistent with a platelike habit. The refinement captures both the peak positions and the relative intensities with a low residual, indicating that the structural model (including texture) describes the experimental data well. We note that the crystallographic information file for $MA_2CuBr_4$ was not publicly available in any of the databases that we have searched (PDF-5+, CCDC, ICSD); therefore, we generated a working structural model from the chloride analogue, as described in Appendix S.E.

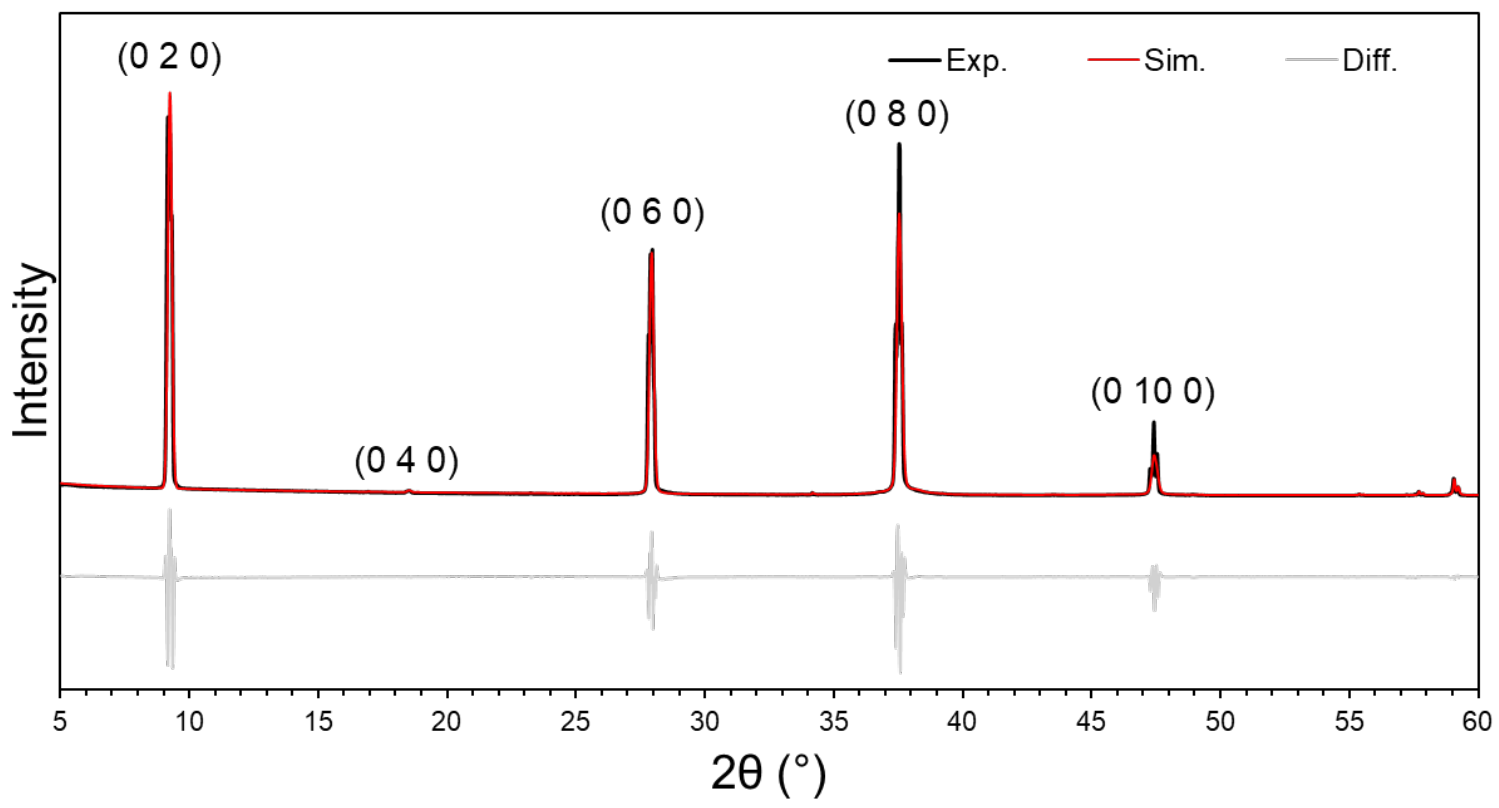

**Figure S.B7**. $MA_2CuBr_4$ (Rietveld refinement). Experimental (black) and calculated (red) powder XRD patterns for $MA_2CuBr_4$ with the difference curve shown in gray. Preferred orientation is modeled using a March–Dollase parameter of 0.21 along [010], consistent with a platelike morphology. The Rietveld refinement converged to $R_{wp}$ = 12.66% and $\chi^2$ = 1.89.

**Pb:**

$MAPbBr_3$ powder diffraction pattern (Figure S.B8) is consistent with a single-phase sample of methylammonium lead bromide. The fit indicates a pronounced texture effect, which is well captured by a March–Dollase preferred-orientation correction (MD = 0.26 along [100]), consistent with a platelike crystallite morphology aligned in this direction. Overall, the calculated profile reproduces the measured peak positions and intensities closely, and the difference curve remains low, supporting the adequacy of the structural model and the absence of major secondary phases.

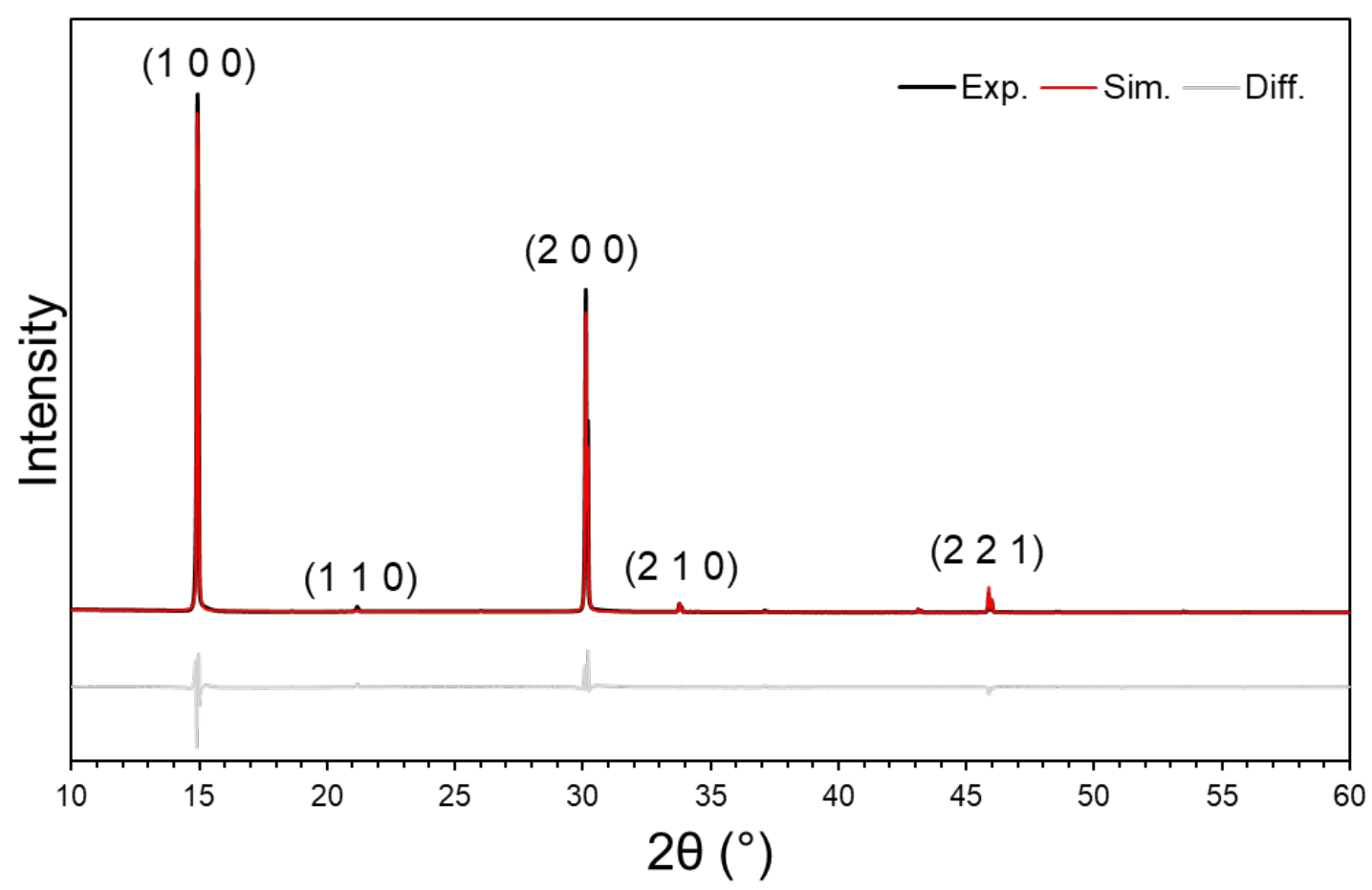

**Figure S.B8**. $MAPbBr_3$ (Rietveld refinement). Experimental (black) and calculated (red) powder XRD patterns for $MAPbBr_3$ with the difference curve shown in gray. The sample exhibits a single phase with preferred orientation described by a March–Dollase parameter of 0.26 along [100] (platelike habit). The Rietveld refinement converged to $R_{wp}$ = 11.12% and $\chi^2$ = 1.82.

**Sn:**

For the Sn-Br system, the powder pattern (Figure S.B9) is best described by a two-phase mixture comprising $MA_2SnBr_6$ and $MASnBr_3$, with $MASnBr_3$ indexed in the $Pmc2_1$ setting. The phase fractions obtained from the fit are 19.9 wt% $MA_2SnBr_6$ and 80.1 wt% $MASnBr_3$ ($Pmc2_1$). The calculated profile accounts for the full set of observed reflections from both phases, and the residual remains low overall, supporting the mixed-phase assignment and the reported quantitative phase composition. When using a milder $Br_2$-reduced PBM solution, namely 1:0.25:1 ratio between MABr:$Br_2$:HBr, the powder diffraction pattern (Figure S.B10) is well described by a two-phase model comprising $MASnBr_3$ as the major phase and metallic Sn as a secondary phase. Rietveld refinement yielded phase fractions of 75.3 wt.% $MASnBr_3$ and 24.7 wt.% Sn ("white tin"). Preferred orientation effects were accounted for using March-Dollase corrections along [100], with refined MD parameters of 0.23 for $MASnBr_3$ and 0.20 for Sn, consistent with platelike crystallite morphology. The calculated pattern reproduces the measured diffraction data well, and the small difference curve indicates that the refinement adequately captures the observed phase composition and texture effects. XPS analysis (Figure S.B11) shows that in both cases surface oxidation occurs (i.e., a $Sn^{4+}$-related peak is present), but between the two reacted films, the one reacted with the milder PBM had a significantly smaller contribution from $Sn^{4+}$.

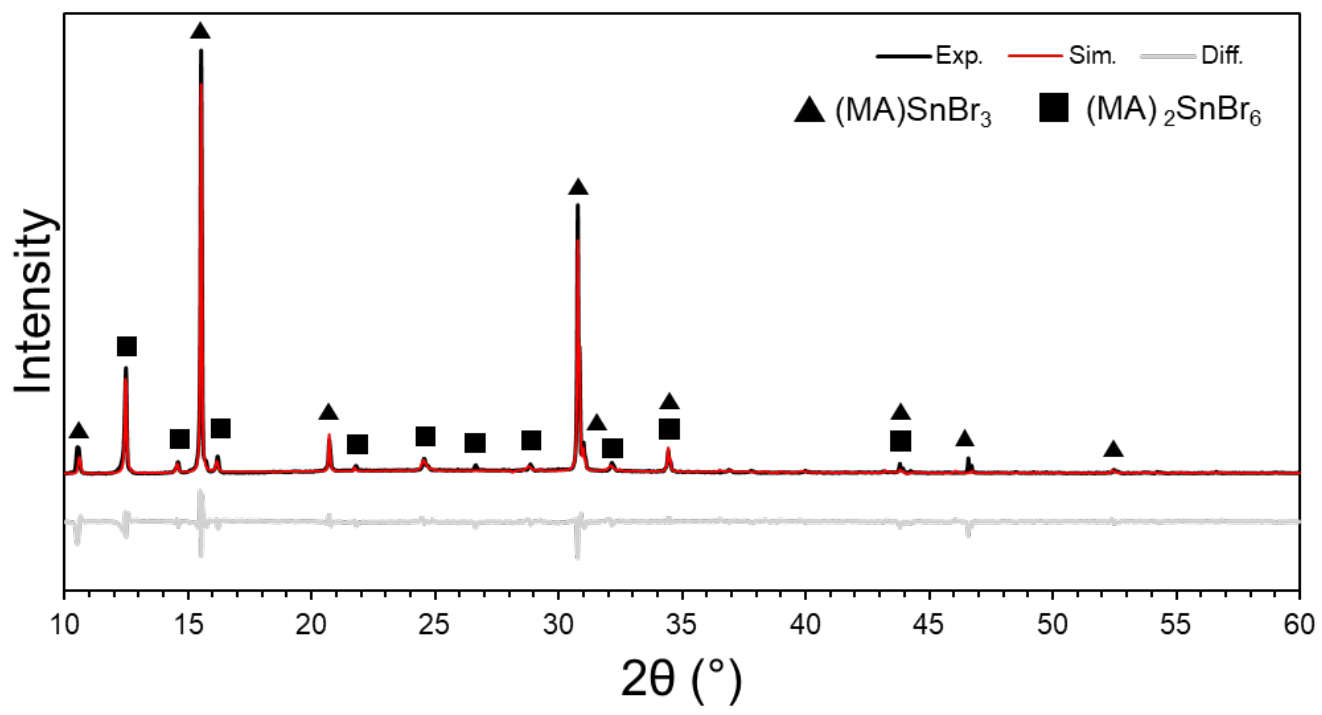

**Figure S.B9:** Sn-Br system (two-phase Rietveld refinement). Experimental (black) and calculated (red) powder XRD patterns for the MA-Sn-Br sample, refined as a two-phase mixture of $MA_2SnBr_6$ (19.9 wt%) and $MASnBr_3$ ($Pmc2_1$; 80.1 wt%), with the difference curve shown in gray. The Rietveld refinement converged to $R_{wp}$ = 13.61% and $\chi^2$ = 1.95.

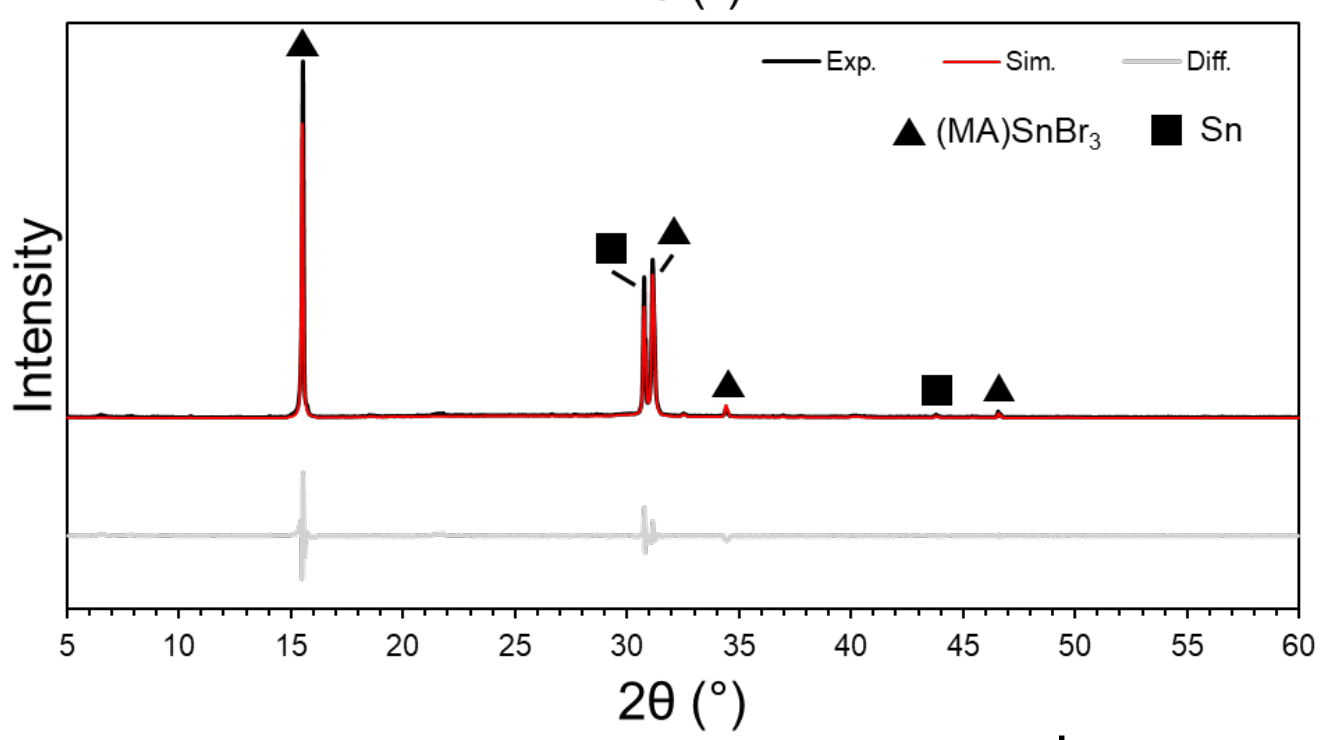

**Figure S.B10**: $MASnBr_3$ Rietveld refinement, showing the observed, calculated, and difference profiles. Quantitative phase analysis revealed $MASnBr_3$ as the major phase (75.3 wt.%) together with metallic Sn ("white tin", 24.7 wt.%). Preferred orientation was accounted for using March-Dollase corrections along [1 0 0], with refined MD parameters of 0.23 for $MASnBr_3$ and 0.20 for Sn, indicating platelike preferred orientation. The refinement converged to $R_{wp}$ = 12.69% and $\chi^2$ = 1.91.

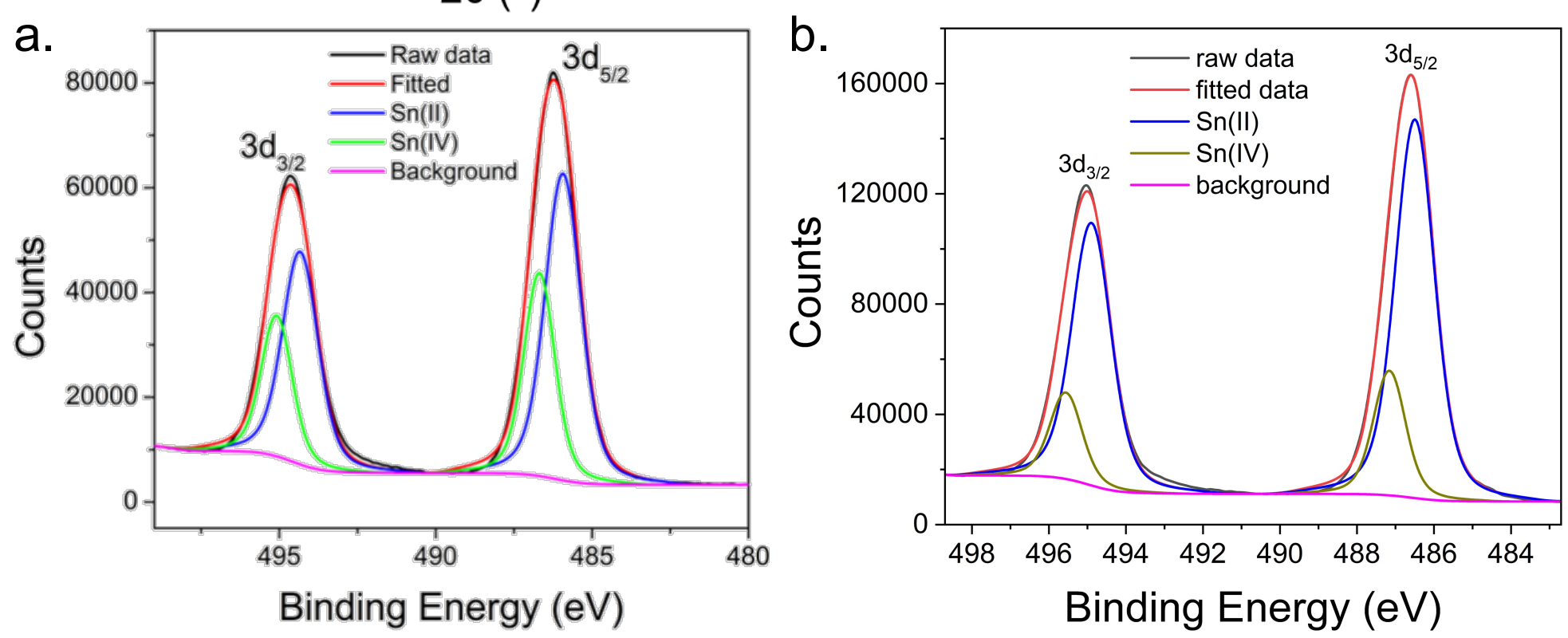

**Figure S.B11**: XPS analysis of Sn metal conversion using (a) the default PBM (1:1:0.1) and (b) the modified PBM ($Br_2$-reduced – 1:0.25:1). The surfaces were analysed after conversion by deposition of PBM on a spin-coating metal film. The molar ratios refer to MABr:$Br_2$:HBr. The calculated Sn(2+):Sn(4+) ratios in the default and modified PBM are 64:36 and 79:21, respectively. No Sn(0) was detected on the analysed surface.

# Appendix S.C: Metal:PBM *in situ* reaction: Supplementary Material

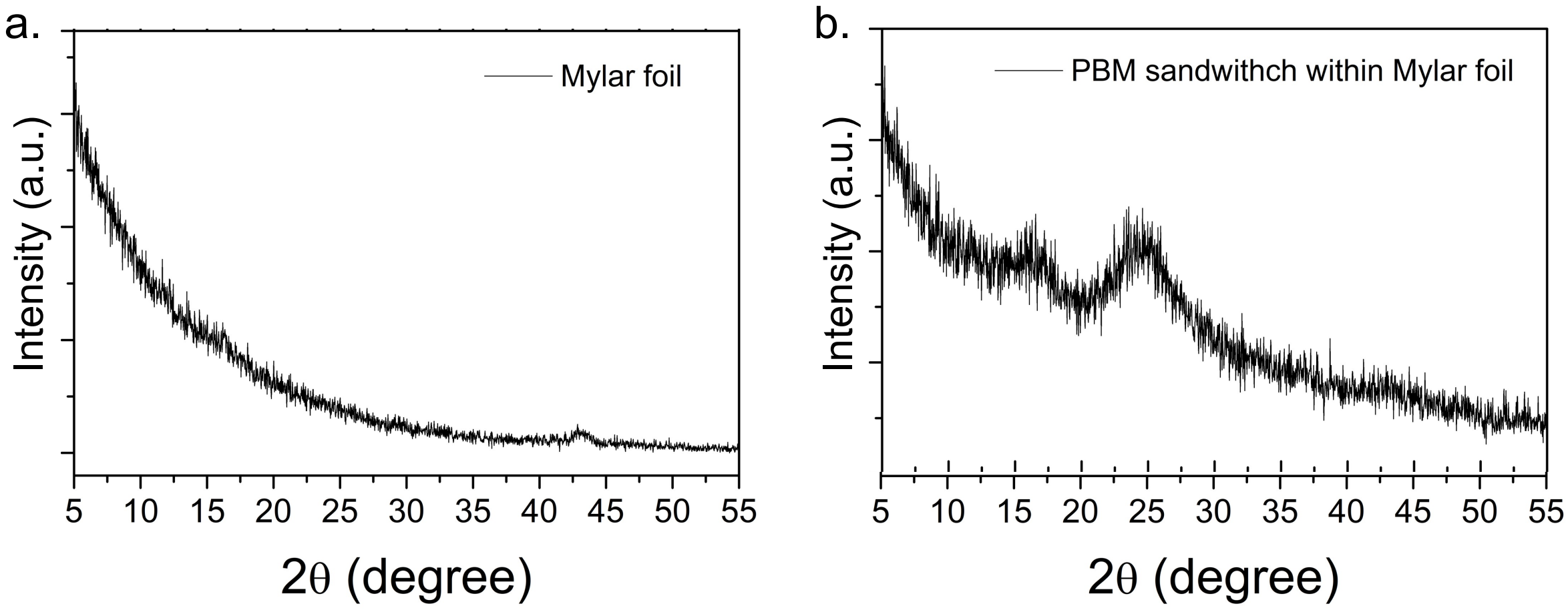

**Figure S.C1**:Transmission mode XRD measurements of the *in situ* reactor background: (a) Mylar foil (*bottom*), and (b) the PBM solution sandwiched within Mylar foil.

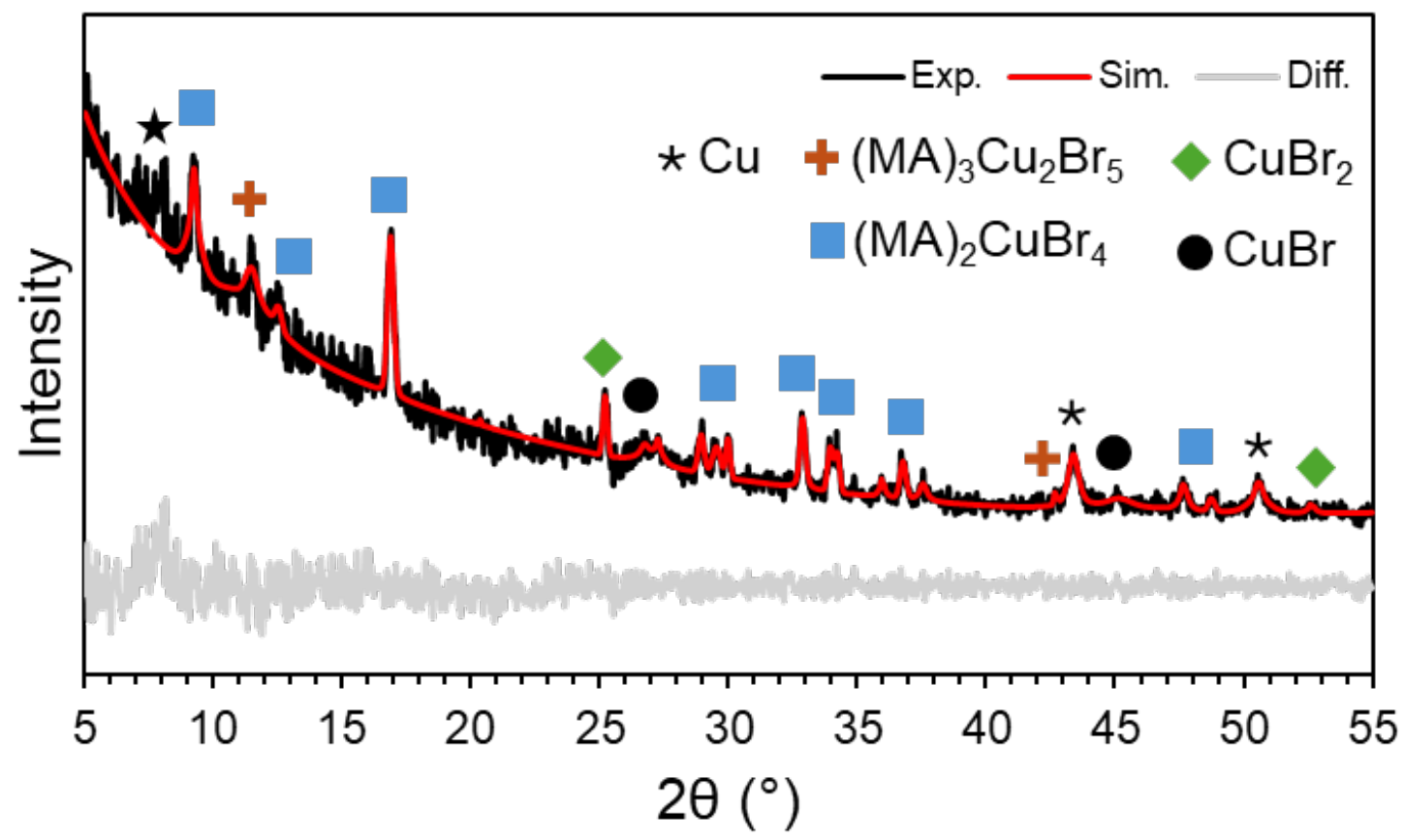

**Figure S.C2.** Rietveld refinement of reacted Cu with PBM after 15 min (see Figure 4, main text). The refinement converged to $R_{wp}$ = 7.35% and $\chi^2$ = 1.63. The reflection at ca. 2θ = 7 (denoted by a black star) corresponds to a large spacing of approximately 12.6 Å, which is characteristic of an expanded low-dimensional or lamellar intermediate rather than a simple Cu-Br salt. Similar transient solvate/intermediate phases are well known in hybrid methylammonium halide crystallization, where precursor composition and solvent coordination strongly affect the crystallization pathway.[13,14] The short lifetime of this reflection suggests formation of a metastable MA/Cu/Br-rich solvated halometallate intermediate that subsequently reorganizes into the denser final crystalline phase. The $(MA)_3Cu_2Br_5$ phase also forms initially as an intermediate phase, but is quickly replaced by $(MA)_2CuBr_4$ as the dominant phase.

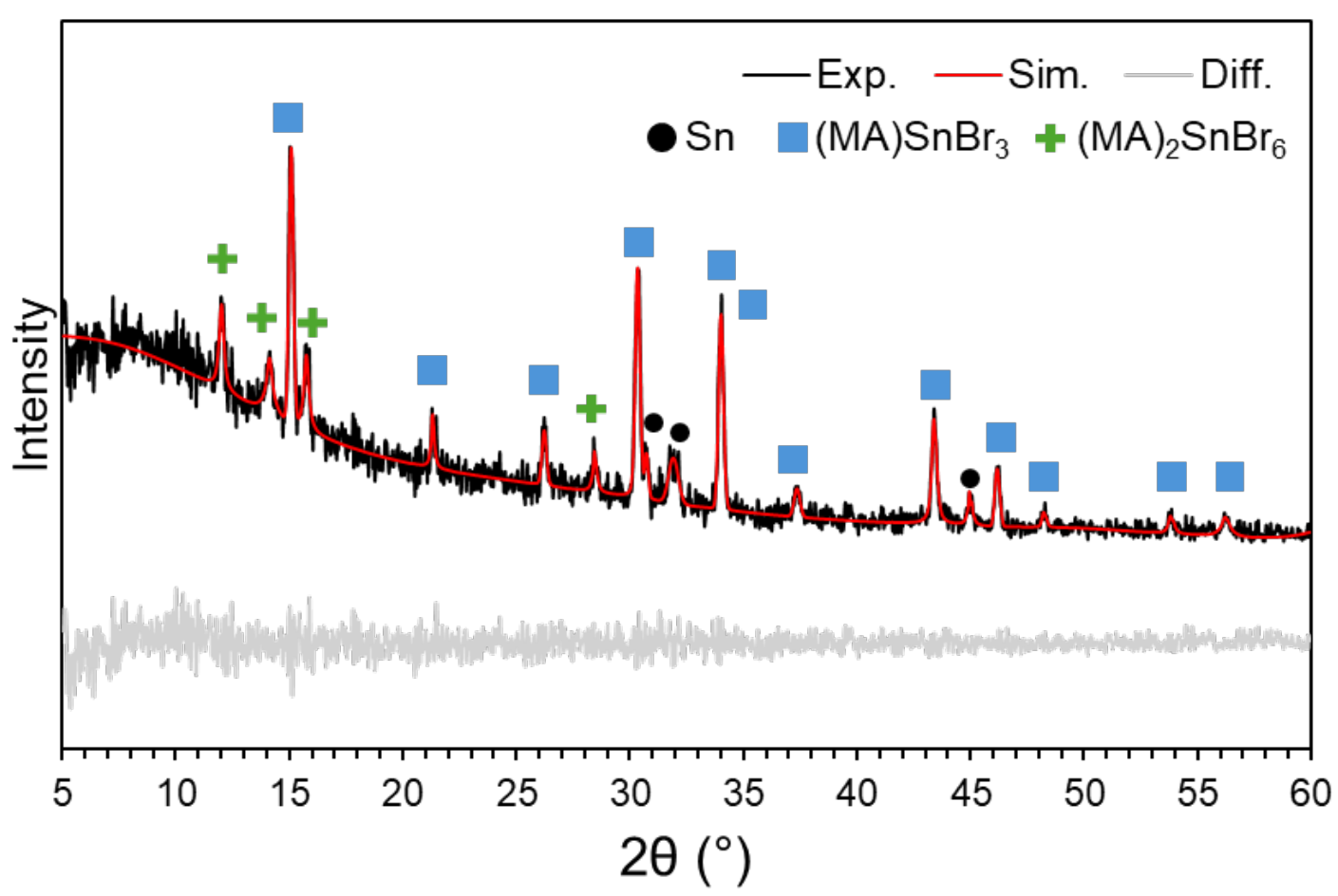

**Figure S.C3.** Rietveld refinement of reacted Sn with PBM after 15 min (see Figure 4, main text). The refinement showed a multi-phase distribution of $(MA)SnBr_3$ (67.9% wt.), $(MA)_2SnBr_6$ (27.3% wt.), and Sn (4.8% wt.). The refinement converged to $R_{wp}$ = 8.71% and $\chi^2$ = 1.93.

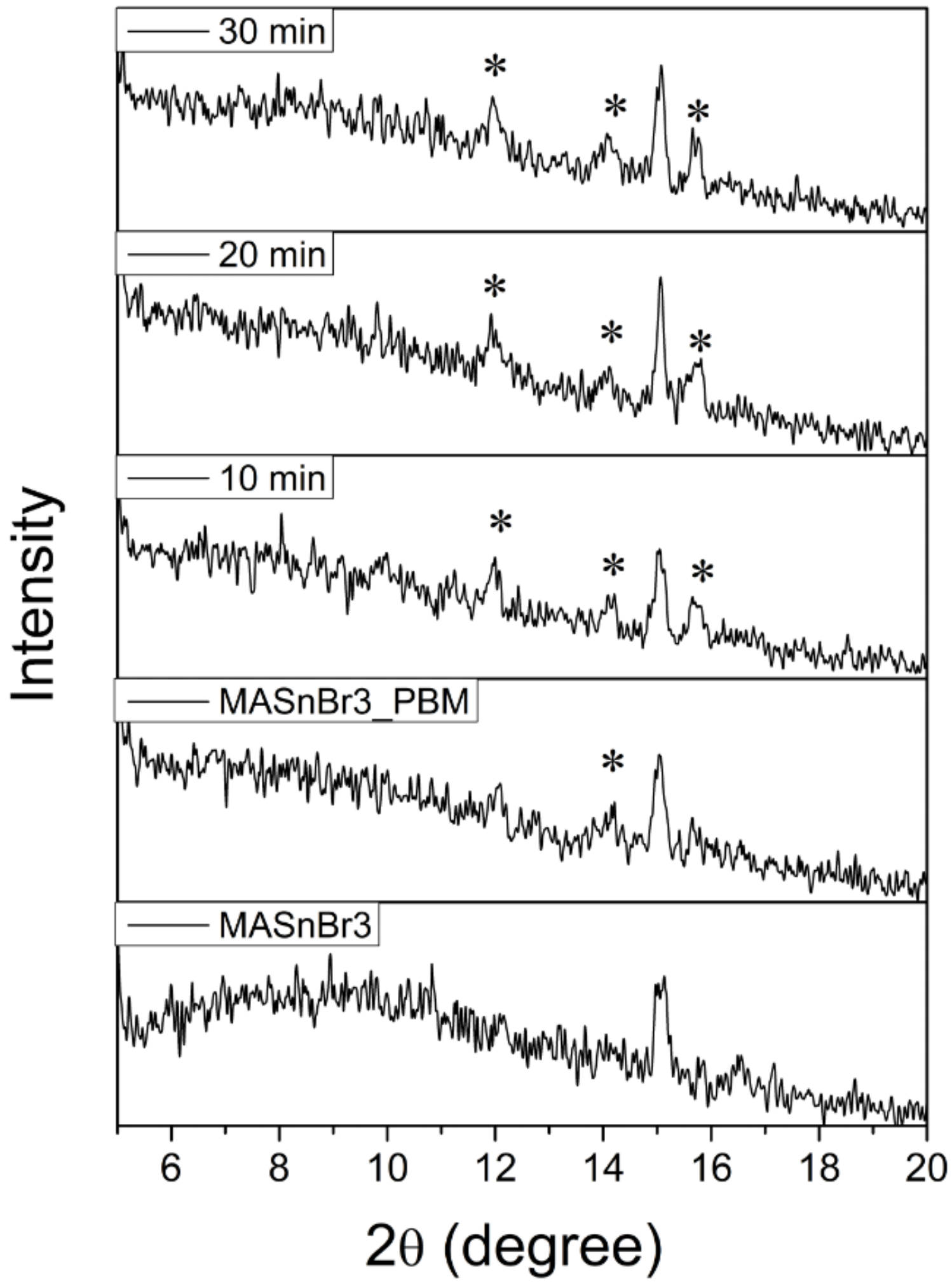

**Figure S.C4**: *in situ* XRD experiment results showing the reaction between $MASnBr_3$ and the default PBM (1:1:0.1) over time. The (*) symbols represent the formation of the $MA_2SnBr_6$ phase (see green "+" sign in Figure S.C3).

# Appendix S.D: *M*+PBM Grinding Experiment: Supplementary Material

## Powder XRD + Rietveld refinement analysis

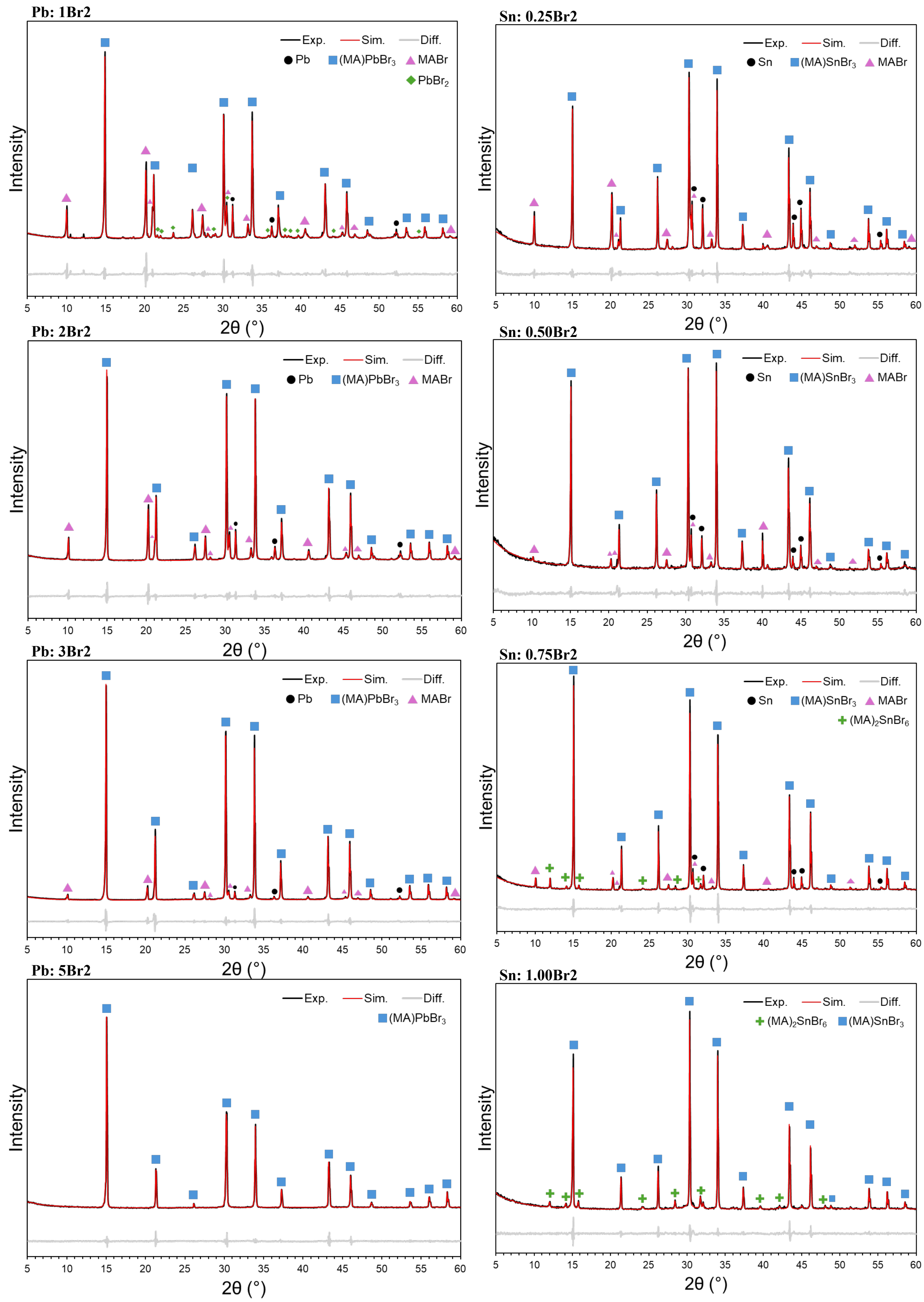

**Figure S.D1**: Powder XRD of patterns and their Rietveld refinement after grinding in a mortar and pestle **Pb** and **Sn** powders with different compositions of PBM. The detailed compositions and phase fractions from the Rietveld analysis are presented in Table S.D1.

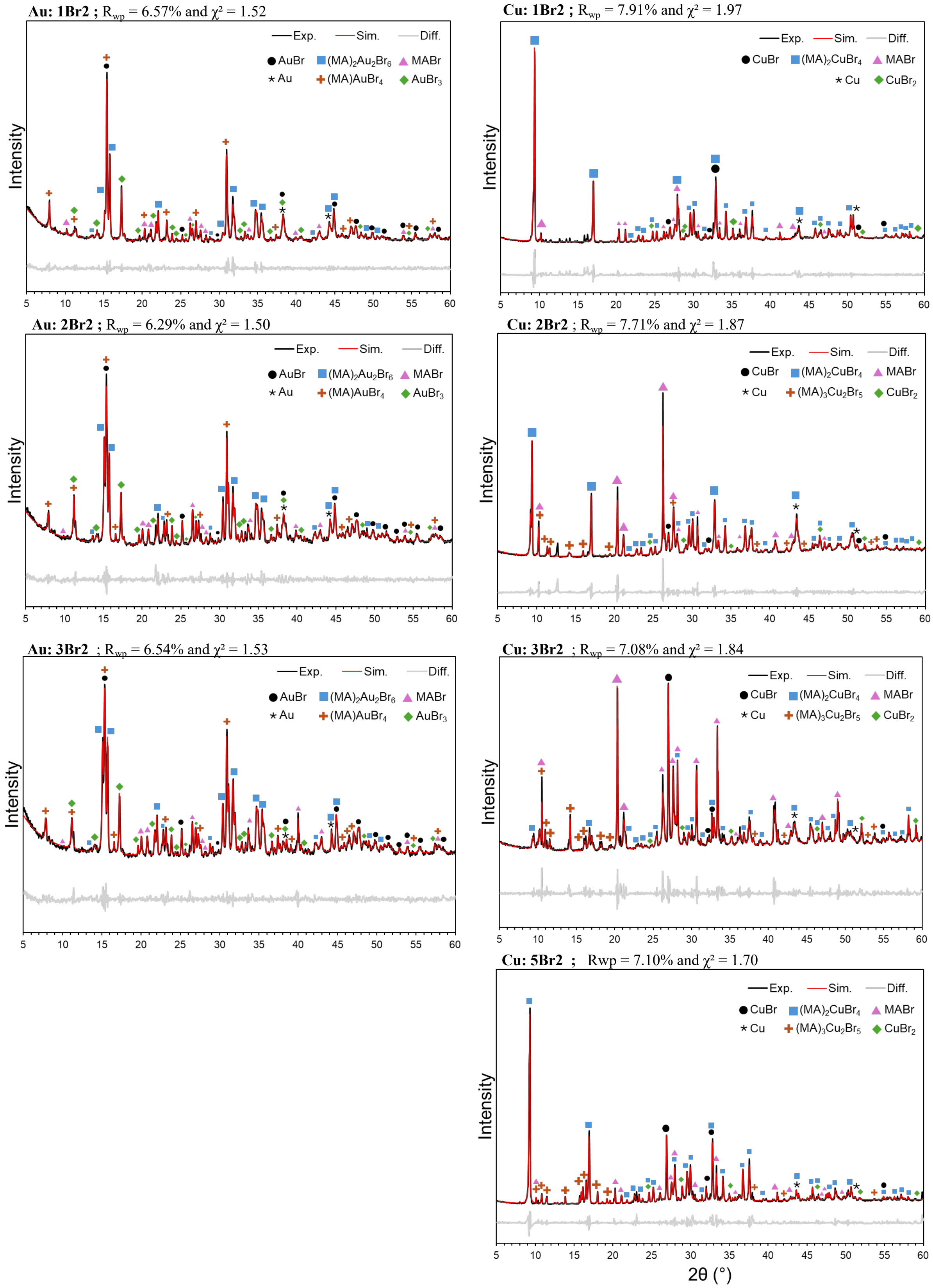

**Figure S.D2**: Powder XRD of patterns and their Rietveld refinement after grinding **Au** and **Cu** powders with different PBM compositions in a mortar and pestle. The detailed compositions and phase fractions from the Rietveld analysis are presented in Table S.D1.

**Table S.D1**: Compositional quantification in mol%, which are derived from wt.% following the Rietveld refinements (Figures S.E1-2) from the ground Metal+PBM products, with varying PBM compositions. In all cases, the [Metal:MABr] mole ratio was 1:1.

| Sample name | PBM content [MABr: $Br_2$:HBr] mol-ratio | Composition (mol%) | | | | | |
|---|---|---|---|---|---|---|---|
| **Pb** | | **$MAPbBr_3$** | **$PbBr_2$** | **MABr** | **Pb** | | |
| **Pb : 1$Br_2$** | **1:1:0.1** | 48.6 | 1.7 | 39.4 | 10.3 | | |
| **Pb : 2$Br_2$** | **1:2:0.1** | 57.4 | ---- | 31.5 | 11.1 | | |
| **Pb : 3$Br_2$** | **1:3:0.1** | 79.5 | ---- | 16.0 | 4.5 | | |
| **Pb : 5$Br_2$** | **1:5:0.1** | 100.0 | ---- | ---- | ---- | | |
| **Sn** | | **$MASnBr_3$** | **$(MA)_2SnBr_6$** | **MABr** | **Sn** | | |
| **Sn : 0.25$Br_2$** | **1:0.25:1** | 49.2 | ---- | 29.5 | 21.3 | | |
| **Sn : 0.50$Br_2$** | **1:0.50:1** | 70.7 | ---- | 8.9 | 20.4 | | |
| **Sn : 0.75$Br_2$** | **1:075:1** | 76.2 | 3.8 | 10.5 | 9.6 | | |
| **Sn : 1.00$Br_2$** | **1:1.00:1** | 90.9 | 9.1 | ---- | ---- | | |
| **Au** | | **$(MA)_2Au_2Br_6$** | **$(MA)AuBr_4$** | **MABr** | **$AuBr_3$** | **AuBr** | **Au** |
| **Au : 1$Br_2$** | **1:1:0.1** | 15.7 | 22.2 | 29.3 | 14.2 | 13.5 | 5.0 |
| **Au : 2$Br_2$** | **1:2:0.1** | 18.9 | 31.7 | 18.9 | 15.0 | 10.7 | 4.8 |
| **Au : 3$Br_2$** | **1:3:0.1** | 15.6 | 42.5 | 18.5 | 11.5 | 10.1 | 1.8 |
| **Cu** | | **$(MA)_2CuBr_4$** | **$(MA)_3Cu_2Br_5$** | **MABr** | **$CuBr_2$** | **CuBr** | **Cu** |
| **Cu : 1$Br_2$** | **1:1:0.1** | 28.7 | --- | 45.0 | 9.8 | 10.2 | 6.4 |
| **Cu : 2$Br_2$** | **1:2:0.1** | 18.5 | 5.7 | 59.3 | 6.4 | 5.6 | 4.5 |
| **Cu : 3$Br_2$** | **1:3:0.1** | 5.9 | 15.6 | 59.3 | 3.2 | 13.1 | 3.0 |
| **Cu : 5$Br_2$** | **1:5:0.1** | 33.8 | 16.7 | 26.5 | 7.4 | 13.9 | 1.7 |

**Table S.D2.** Crystallographic parameters and structural references for the perovskite and perovskite-related phases used in Rietveld refinement.

| Phase | Space group | a, b, c (Å) | β (°) | Structure Determination Method |
|---|---|---|---|---|
| **$MAPbBr_3$** | *$Pm\bar{3}m$*, No. 221 | 5.9259 (2) | 90 | Ref. [15] |
| **$MASnBr_3$** | *$Pmc2_1$, No. 26* | 5.8941 (2), 8.3862 (2), 8.2406 (2) | 90 | Lower-symmetry $MASnBr_3$ structure reported by Swainson et al. – Ref. [16]. Used here because this model reproduced the observed $MASnBr_3$ reflections, whereas the cubic *$Pm\bar{3}m$* model did not. |
| **$MA_2CuBr_4$** | *Pbca*, No. 61 | 7.5777 (4), 19.0890 (9), 7.8301 (4) | 90 | Generated from the layered chloride analogue, $MA_2CuCl_4$, by Cl to Br substitution, followed by refinement against the experimental PXRD data. The chloride / bromide structural relationship follows Ref. [17]. For more details, see Appendix S.E. |
| **$MA_2Au_2Br_6$** | *$P4_12_12$*, No. 92 | 7.9115(10), 23.385(3) | 90 | Ref. [7] |
| **$MAAuBr_4$** | *C2/c*, No. 15 | 32.738(4), 8.4616(11), 7.2142(9) | 102.134 (2) | [a] Generated from the reported anhydrous $MAAuCl_4$ structure [18] by Cl to Br substitution, followed by refinement of the unit-cell parameters and structural model against the experimental PXRD pattern. For more details, see Appendix S.E. |
| **$MA_2SnBr_6$** | *$R\bar{3}m$*, No. 166 | 7.5334 (5), 7.5334 (5), 22.1364 (14) | 90 | Structure determined by single-crystal XRD from laboratory-grown crystals – Generated structure: CCDC 2512222. For more details, see Appendix S.E. |
| **$MA_3Cu_2Br_5$** | *Pnma*, No. 62 | 9.6831 (3), 11.0157 (5), 13.6225 (4) | 90 | Generated model based on the $Cs_3Cu_2Br_5$-type structure,[11] with $MA^+$ introduced in place of $Cs^+$. |

## Appendix S.E: Structure determination of $MA_2CuBr_4$, $MAAuBr_4$, $MA_2SnBr_6$

For Rietveld refinement, we needed crystal structures with atomic positions. While for most analyzed phases these were present in the PDF5+ database, some phases were absent, in particular $MA_2CuBr_4$, $MAAuBr_4$, and $MA_2SnBr_6$. Therefore, we had to generate these for our analysis throughout this work. Below, we specify the approaches we used for generating their crystallographic phases for analysis.

$MA_2CuBr_4$:

The initial structural model for $(MA)_2CuBr_4$ was constructed by using the known layered copper(II) chloride analogue, $(MA)_2CuCl_4$, as a chemically reasonable prototype. In this model, the Cl atoms were replaced by Br atoms, while preserving the connectivity of the two-dimensional $CuX_6$ framework and the methylammonium sublattice. This approach is justified by the known structural relationship within the $(MA)_2Cu(Cl_xBr_{1-x})_4$ family: both end members consist of quasi-two-dimensional $Cu^{2+}$ halide layers separated by methylammonium cations, and the chloride and bromide compounds were already reported by Kimishima to share the same general layered structural type. In particular, the 1980 study lists both $(MA)_2CuCl_4$ and $(MA)_2CuBr_4$ in the orthorhombic *Pbca* setting, with the bromide analog showing the expected expansion of the lattice parameters from a = 7.535, b = 7.30, c = 18.60 Å for the chloride to a = 7.820, b = 7.64, c = 19.18 Å for the bromide.[17] The substituted model was then used only as a starting point for Rietveld refinement; the unit-cell parameters, relative atomic coordinates, and profile parameters were refined against the experimental PXRD pattern until convergence. Thus, the final bromide structure was determined by the diffraction data and was not imposed by the chloride prototype.

$MAAuBr_4$:

As an independent crystal structure of anhydrous $MAAuBr_4$ was not available, the phase was modeled using the reported $MAAuCl_4$ structure as a crystallographic prototype. The chloride sites were substituted by bromide, after which the unit-cell parameters and structural model were refined against the experimental PXRD pattern using Rietveld refinement. This approach produced an excellent agreement with the observed diffraction pattern, supporting the assignment of the dominant phase as an anhydrous methylammonium tetrabromoaurate phase.[18]

$MA_2SnBr_6$:

To generate the $MA_2SnBr_6$ structural model, single crystals were grown following ref.[19] that was originally proposed to grow $MASnBr_3$ crystals. The precursor solution contained 0.674 g SnO (5 mmol, molar mass 134.721 gm/mol) and 0.795 g MABr (7.1 mmol, molar mass 111.97 gm/mol) dissolved in 12 mL HBr and 6 mL hypophosphorous acid ($H_3PO_2$). Solutions were stirred at 80 °C until fully dissolved, filtered, and then heated at 85 °C for 24 h. Crystallization commenced after ~4 h, yielding small red-brown grains mixed with yellow particles. We managed to isolate the pale yellow crystals, which were verified as $MA_2SnBr_6$. The $MASnBr_3$ phase was used for other experiments as described in the text (resulting in Figure S.C4).

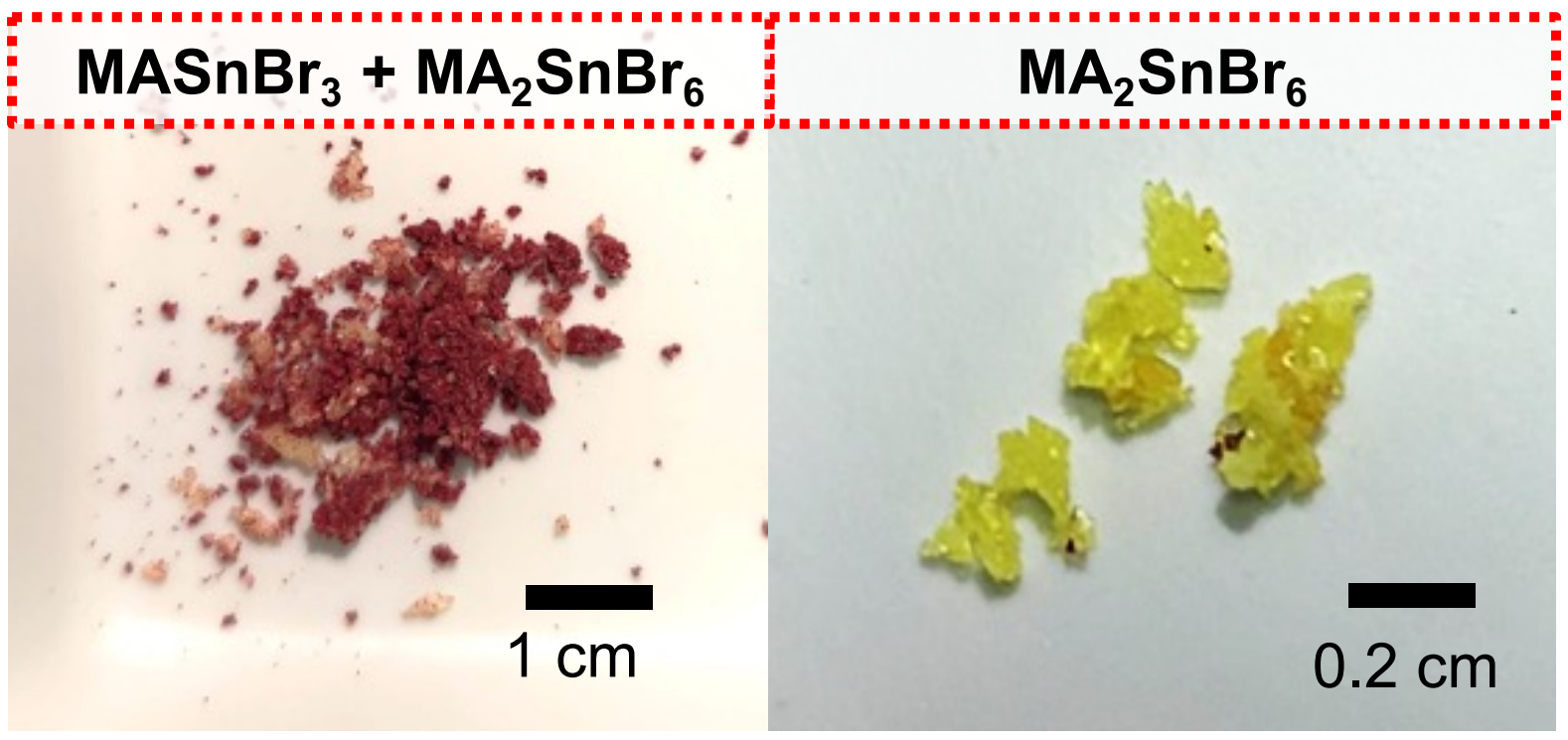

**Figure S.E1:** $MASnBr_3$ and $MA_2SnBr_6$ single crystals obtained using the growth protocol described above.

## $MA_2SnBr_6$ single crystal XRD and its structural refinement

The crystal was mounted on a CrystalCap™ ALS HT cryo-loop mount for data collection on a Rigaku XtaLAB Synergy-S single crystal x-ray diffractometer, which includes a HyPix-Arc 100° detector and a standard Cu *Kα* x-ray radiation source (λ = 1.54184 Å). Unit cell dimensions, space group assignment, data reduction, and finalization were done by using the CrysAlis PRO software package[20] (ver. 171.43, released 2024).

**$MA_2SnBr_6$:** A pale-yellow plate-like crystal was used for analysis. A total of 1356 reflections were collected, of which 1155 were used. The structure was solved in the trigonal *R-3m* space group (no. 166) via intrinsic phasing and refined by SHELXL using a full-matrix least-squares technique. The structure is shown in Figure S.E2, and a summary of crystal data is given in Table S.E1.

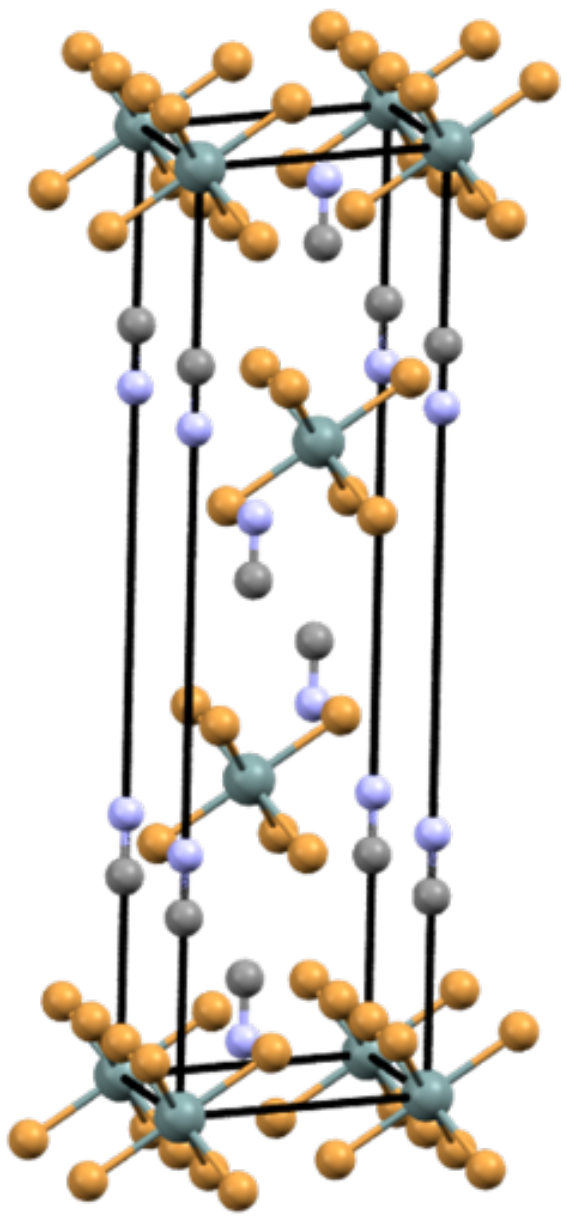

**Figure S.E2:** Crystal structures of $MA_2SnBr_6$. Color code: yellow spheres - Br, grey spheres - C, blue spheres – N, teal spheres - Sn. Hydrogens are omitted for clarity. The crystal structure of $MA_2SnBr_6$ is trigonal (rhombohedral), space group *R-3m*. The inorganic sublattice consists of isolated $[SnBr_6]^{2-}$ octahedra, in which each $Sn^{4+}$ center is octahedrally coordinated by six bromide anions with no corner-, edge- or face-sharing between neighbouring octahedra. The methylammonium ($MA^+$) cations occupy the voids around these $[SnBr_6]^{2-}$ units, effectively surrounding each octahedron and separating it from the next. As a consequence, there is no continuous Sn-Br framework or sheet formation; instead, $MA_2SnBr_6$ is best described as a zero-dimensional ($A_2BX_6$-type) vacancy-ordered perovskite composed of discrete, molecular-like $[SnBr_6]^{2-}$ entities embedded in an $MA^+$ matrix. Summary of crystal data for the material is given in Table S.E1. The corresponding structure is uploaded to CCDC under the number 2512222; the CIF is provided as a separate attachment.

**Table S.E1:** Crystal data and structure refinement details of $MA_2SnBr_6$.

| Material | $MA_2SnBr_6$ |
|---|---|
| **Chemical Formula** | $(H_3CNH_3)_2SnBr_6$ |
| **$M_r$ (g·mol$^{-1}$)** | 662.29 |
| **Crystal system, space group** | Rhombohedral, *R-3m* |
| ***a, c* (Å)** | 7.5334(5), 22.1364(14) |
| ***V* (Å$^3$)** | 1087.98(16) |
| ***Z*** | 3 |
| **$\rho_{calcd}$ (g·cm$^{-3}$)** | 3.032 |
| **CCDC number** | 2512222 |

## Appendix S.F: Raman spectroscopy of the reacted metals

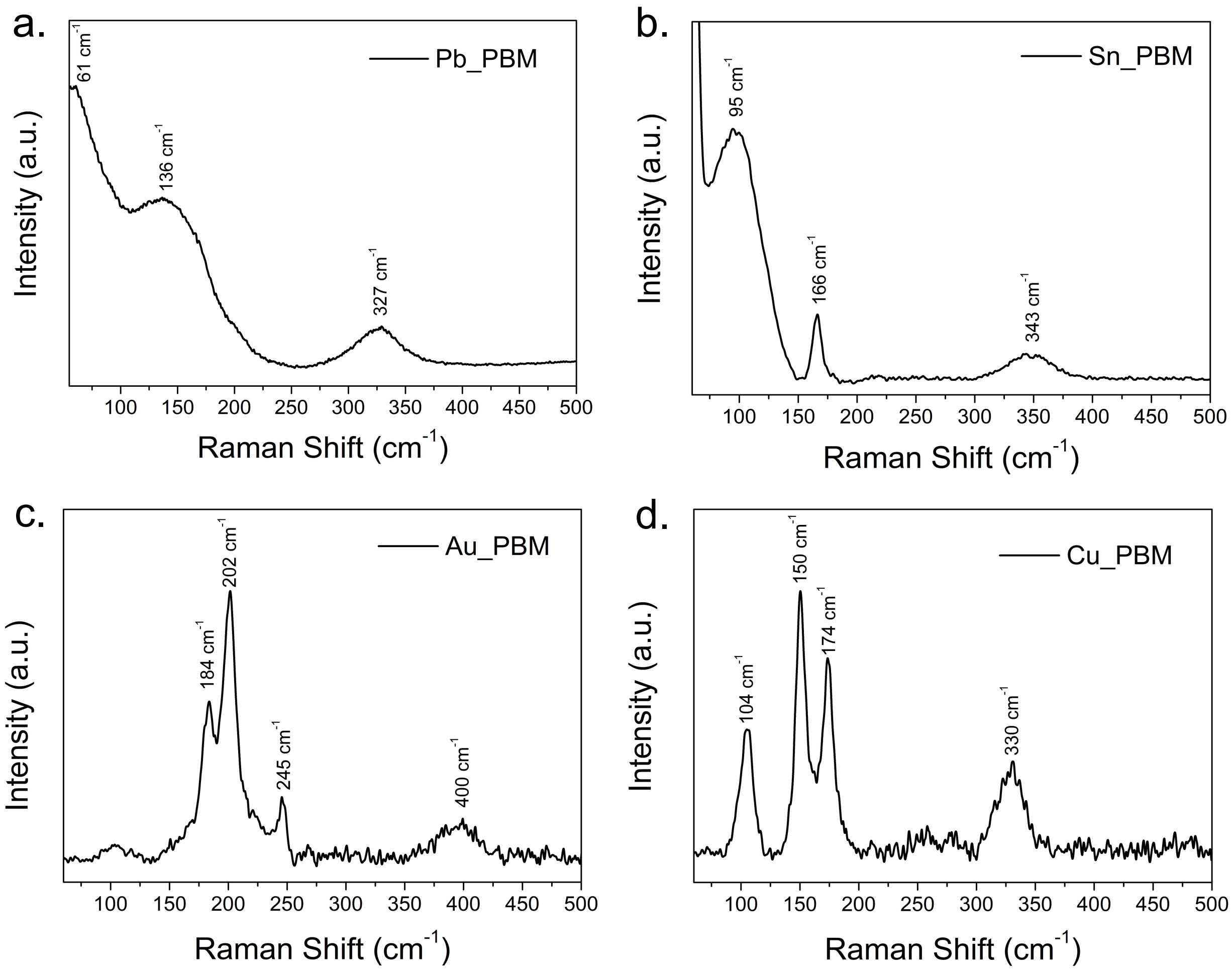

**Figure S.F1:** Raman spectra that follow the M+PBM reaction using our most optimized chemical composition of the PBM that resulted in the purest form of the metal-perovskite. Specifically: for Pb and Sn we used the grinding experiment with a 1:5:0.1 and 1:1:1 PBM, respectively; for Au and Cu we used spattered films followed by spin-coating of a 1:1:0.1 PBM; the PBM mole ratio correspond to MABr:$Br_2$:HBr. The peak values are comparable to those published in literature: $MAPbBr_3$ [21–23] ; $MASnBr_3$ [24,25] ; $MA_2Au_2Br_6$ [26,27] ; $MA_2CuBr_4$ [28]. *M*–Br vibrations are observed up to 300 $cm^{-1}$, while bands in the range of 300 – 450 $cm^{-1}$ are associated with the restricted rotation of the $MA^+$ cation.

## Appendix S.G: Methods: Materials and Analysis Details

### Usage of AI-generative tools

We disclose that generative AI tools were not used except for the TOC figure and the illustrations in Figure 2a (main text). They were generated using ChatGPT and Gemini image-generation tools, combined with original editing. They were used only for clarity and illustration purposes; they did not use data, nor were they essential for the scientific part of the manuscript.

### Precursor Materials

Methylammonium bromide (Ossilla, ≥98%), liquid bromine (Sigma Aldrich, ≥99.9%, density = 3.11 $gcm^{-3}$ at 25°C), hydrobromic acid (HBr, 48 wt%), metallic lead sputtering target, metallic lead powder (100 mesh, 99.95%, Thermo Scientific), metallic Cu powder (150 mesh, Thermo Scientific, 99.95%), Au coil (99.999%), metallic Au powder scratched for Au evaporated foil, metallic Sn powder (325 mesh, 99.8% Alfa Aesar), isopropanol (analytical reagent>99.5%), acetone (analytical reagent>99.8%), ethanol (analytical reagent>99.5%), diethyl ether (analytical reagent>99.8%). Since the main synthesis process is an essential part of the novelty of this paper, we will describe it in detail in the *Results* section.

### Metallic Thin Film Deposition

For the fabrication of metallic films, clean 1×1 $cm^2$ glass substrates were used. Glass slides were cut into 1×1 $cm^2$ parts, then cleaned sequentially with Alconox powder and de-ionized (DI) water, and then sonicated in DI water, acetone, and isopropanol for 10 min each followed by 5 min of UV-ozone treatment. Metallic films were deposited using magnetron sputtering (Pb-200 nm, Sn-500 nm), thermal evaporation (Au-50 nm), or E-gun Sputtering (Cu-200 nm). The identity of the evaporated films was verified via XRD and their thickness via a profilometer. The films were then used for reaction with PBM via spin-coating on the metallic films.

### Reaction of Metallic thin films with PBM

The synthesized PBM was spin-coated at 5000 rpm for 20s on top of the metallic films. The PBM melt is highly viscous/dense, and to minimize the leftover of the unreacted MA salt, we dispensed the PBM 5s after spin coating started. We observed color change in all the metallic films as it reacted with the polybromide melt solutions. After spin coating, the films were annealed at 110°C for 10 min to promote complete conversion and removal of residual volatile components.

### Reaction of Metallic powder with PBM

The reactions of powdered metallic precursors – Pb, Sn, Au, and Cu – with the polybromide melt (PBM) were investigated using a solvent-free grinding approach in a mortar and pestle under varying bromine concentrations. For the Pb-based system, the $Br_2$ concentration was systematically increased up to 500 mol% with respect to the metal powder and MABr. In the case of Sn powder, the $Br_2$ concentration was varied from 25 mol% to 100 mol%. To mitigate the undesired oxidation of $Sn^{2+}$ to $Sn^{4+}$ and stabilize the desired perovskite phase, 100 mol% HBr (with respect to MABr) was incorporated into the reaction mixture. Similarly, for Au powder, $Br_2$ concentration was varied up to 300 mol%. For Cu powder, the $Br_2$ concentration was varied up to 500 mol%.

### Powder X-ray Diffraction (XRD)

Powder X-ray diffraction (PXRD) data were collected on a Panalytical Empyrean diffractometer equipped with a Cu K$\alpha$ X-ray source (λ = 1.5418 Å). Diffraction patterns were analyzed by Rietveld refinement using the GSAS-II software package (version 5711, 2024 release)[29], or HighScore plus version 5.2, which is coupled with the Panalytical Empyrean diffractometer, in the cases of the *in situ* experiments. Structural models of the main and secondary phases were extracted from the International Center for Diffraction Data (ICDD) Powder Diffraction File (PDF-5+) database (2025). In the cases where the relevant perovskite structure was not available in the database or the literature, it was generated from similar structure files of related chloride analogues/prototypes (in the cases of $MAAuBr_4$ and $MA_2CuBr_4$)[17,18] or measured from a single-crystal XRD measurement using laboratory-grown crystals (the case of $MA_2SnBr_6$) – see Appendix S.E:.

Phase identification was first established using reference diffraction data, followed by full-pattern refinement to determine lattice parameters, relative phase abundances, and sample-dependent structural effects. Background intensities were fitted using a Chebyshev polynomial function, and reflection profiles were modeled with pseudo-Voigt peak-shape functions. Refinements were performed by a full-matrix least-squares procedure, and quantitative phase analysis included a microabsorption correction. For samples exhibiting non-random crystallite orientation, preferred orientation was refined using the March-Dollase model. The final refinement included instrumental parameters, background terms, unit-cell parameters, scale factors, profile-shape coefficients, phase fractions, and structural parameters as applicable.

*Powder XRD Measurement modes*: We used two powder XRD measurement modes: *reflection* for all ex-situ reacted films and ground powders and *transmission* for *in situ* reaction of metallic powders with PBM, where the reaction took place encapsulated between two Mylar® (polyethylene Terephthalate, PET) foils. For the reflection mode (Bragg-Brentano geometry), the typical step size and dwell time were 0.0336° 2θ and 45 seconds per step, respectively. For the transmission geometry measurement, we used a sandwiched Mylar® system embedded in a spinning sample holder; the reaction chamber is shown in **Figure 4** of the main text. The two Mylar® film layers were clipped together and sealed closed with Teflon™ (polytetrafluoroethylene, PTFE) rings to prevent any reaction between the PBM and the reaction chamber.

For the *in situ XRD experiment*, we tested in both transmission and reflection modes. We found that the transmission mode was more robust and consistent across the Mylar background; thus, we proceeded with it. In every experiment, we started with ~20 mg of each metal, Au, Cu, Pb, and Sn powder, which was sandwiched between two Mylar® films and measured by XRD before the addition of the PBM. The closed Mylar® reaction vessel was carefully opened and ~30 µL of PBM was dropped over the metal powder and immediately re-sealed. Because there was a very small gap between the Mylar® films, and the encapsulation prevented air/moisture from entering the reaction medium, the reaction was relatively controlled with only residual oxygen and water present in the sealed holder. The evolving reaction products were measured at 15 min intervals. The 2θ range was dynamically adjusted to ensure that both the pure metallic and perovskite-related peaks were captured. The step size and dwell time were 0.0128° 2θ and approximately 60 seconds per step, though the dwell time was dynamically adjusted to ensure even (15 min) time-step intervals for each measurement sequence plus wait time. All the measurements were performed at room temperature.

## Raman

Micro Raman measurements were performed with a confocal Horiba LabRam HR Evolution, equipped with a Syncerity-OE CCD and Synapse FIVIS EM-CCD detectors. The ground metal-PBM products were measured under $N_2$ atmosphere and with low laser power, ensuring the stability of the reaction products. The PBM and $Br_2$ were measured using a sealed quartz cuvette. For the powder samples, PBM, Cu-PBM, and Au-PBM, a 532 nm laser was used as the excitation source, with power on the sample ranging from 0.035 to 3.5 mW. For $Br_2$, Pb-PBM, and Sn-PBM perovskites, a 633 nm laser was used with power on the sample of 0.12 mW. A 50 x long-working-distance (LWD) objective (Olympus LMPlanFL-N, NA = 0.5) focused the laser on the sample. The confocal hole was set to 100-120 μm, and a 600 gr/mm grating was used for all samples. The acquisition time varied from 0.2 to 30 seconds. Spectra were acquired and processed (including Bose-Einstein, baseline correction, and averaging) using LabSpec 6 software, version 6.8.2.10.

## X-ray photoelectron spectroscopy (XPS)

To investigate the $Sn^{2+}/Sn^{4+}$ oxidation states, we analyzed spin-coated $MASnBr_3$ thin films using X-ray photoelectron spectroscopy (XPS). We employed ESCALAB photoelectron / Auger electron spectroscopy microprobe (Thermo Fisher Scientific, Massachusetts, USA) in the XPS mode, with Al K$\alpha$ X-ray source and 900 μm spot size. All the scans were obtained in a CAE analyzer mode with a pass energy of 20 eV and a step size of 0.1 eV. To gain a sufficient signal-to-noise ratio, the total number of scan repetitions was 10.

## Optical Transmission and PL spectroscopy

Optical transmission was measured using a Cary 5000 UV-Vis-NIR Spectrophotometer equipped with a tungsten-halogen lamp for the visible range and a deuterium arc for the UV range. PL was performed using a Fluorolog-3 HORIBA Scientific with a 450 watt Xenon arc lamp. Controlled by the Cary WinUV software, a modular Windows-based software, makes it easy to perform powerful analysis and control a number of optional accessories.

## Scanning-Electron Microscope (SEM)

For SEM measurements, we used a Thermo-Fisher Verios 460L field-emission scanning electron microscope (FESEM) with an acceleration voltage of 3-10 keV.